\documentclass{aastex631}

\usepackage{amssymb}
\usepackage{mathrsfs}

\begin{document}

\title{A Phase-Resolved View to the Low-frequency Quasi-periodic Oscillations from the Black Hole Binary MAXI J1820+070}

\correspondingauthor{Qing C. Shui}
\email{shuiqc@ihep.ac.cn}
\correspondingauthor{S. Zhang}
\email{szhang@ihep.ac.cn}
\correspondingauthor{Shuang N. Zhang}
\email{zhangsn@ihep.ac.cn}

\author[0000-0001-5160-3344]{Qing C. Shui}
\affiliation{Key Laboratory of Particle Astrophysics, Institute of High Energy Physics, Chinese Academy of Sciences, 100049, Beijing, China}
\affiliation{University of Chinese Academy of Sciences, Chinese Academy of Sciences, 100049, Beijing, China}

%\author{The HXMT Team}
%\affiliation{Key Laboratory of Particle Astrophysics, Institute of High Energy Physics, Chinese Academy of Sciences, 100049, Beijing, China}
%\affiliation{University of Chinese Academy of Sciences, Chinese Academy of Sciences, 100049, Beijing, China}
%\affiliation{Institut f\"{u}r Astronomie und Astrophysik, Kepler Center for Astro and Particle Physics, Eberhard Karls, Universit\"{a}t, Sand 1, D-72076 T\"{u}bingen, Germany}
\author{S. Zhang}
\affiliation{Key Laboratory of Particle Astrophysics, Institute of High Energy Physics, Chinese Academy of Sciences, 100049, Beijing, China}

\author[0000-0001-5586-1017]{Shuang N. Zhang}
\affiliation{Key Laboratory of Particle Astrophysics, Institute of High Energy Physics, Chinese Academy of Sciences, 100049, Beijing, China}
\affiliation{University of Chinese Academy of Sciences, Chinese Academy of Sciences, 100049, Beijing, China}
%\collaboration{20}{(AAS Journals Data Editors)}

\author{Yu P. Chen}
\affiliation{Key Laboratory of Particle Astrophysics, Institute of High Energy Physics, Chinese Academy of Sciences, 100049, Beijing, China}

\author[0000-0003-3188-9079]{Ling D. Kong}
\affiliation{Institut f\"{u}r Astronomie und Astrophysik, Kepler Center for Astro and Particle Physics, Eberhard Karls, Universit\"{a}t, Sand 1, D-72076 T\"{u}bingen, Germany}

\author[0000-0002-6454-9540]{Peng J. Wang}
\affiliation{Key Laboratory of Particle Astrophysics, Institute of High Energy Physics, Chinese Academy of Sciences, 100049, Beijing, China}
\affiliation{University of Chinese Academy of Sciences, Chinese Academy of Sciences, 100049, Beijing, China}

\author[0000-0002-5554-1088]{Jing Q. Peng}
\affiliation{Key Laboratory of Particle Astrophysics, Institute of High Energy Physics, Chinese Academy of Sciences, 100049, Beijing, China}
\affiliation{University of Chinese Academy of Sciences, Chinese Academy of Sciences, 100049, Beijing, China}

\author{L. Ji}
\affiliation{School of Physics and Astronomy, Sun Yat-Sen University, Zhuhai, 519082, China}

\author{A. Santangelo}
\affiliation{Institut f\"{u}r Astronomie und Astrophysik, Kepler Center for Astro and Particle Physics, Eberhard Karls, Universit\"{a}t, Sand 1, D-72076 T\"{u}bingen, Germany}

\author[0000-0002-0638-088X]{Hong X. Yin}
\affiliation{Shandong Key Laboratory of Optical Astronomy and Solar-Terrestrial Environment, School of Space Science and Physics, Institute of Space Sciences, Shandong University, Weihai, Shandong 264209, China}

\author[0000-0002-9796-2585]{Jin L. Qu}
\affiliation{Key Laboratory of Particle Astrophysics, Institute of High Energy Physics, Chinese Academy of Sciences, 100049, Beijing, China}

\author[0000-0002-2705-4338]{L. Tao}
\affiliation{Key Laboratory of Particle Astrophysics, Institute of High Energy Physics, Chinese Academy of Sciences, 100049, Beijing, China}

\author[0000-0002-2749-6638]{Ming Y. Ge}
\affiliation{Key Laboratory of Particle Astrophysics, Institute of High Energy Physics, Chinese Academy of Sciences, 100049, Beijing, China}

\author{Y. Huang}
\affiliation{Key Laboratory of Particle Astrophysics, Institute of High Energy Physics, Chinese Academy of Sciences, 100049, Beijing, China}

\author{L. Zhang}
\affiliation{Key Laboratory of Particle Astrophysics, Institute of High Energy Physics, Chinese Academy of Sciences, 100049, Beijing, China}

\author[0000-0003-2845-1009]{Hong H. Liu}
\affiliation{Center for Field Theory and Particle Physics and Department of Physics, Fudan University, 200438 Shanghai, People's Republic of China}

\author{P. Zhang}
\affiliation{College of Science, China Three Gorges University, Yichang 443002, China}

\author{W. Yu}
\affiliation{Key Laboratory of Particle Astrophysics, Institute of High Energy Physics, Chinese Academy of Sciences, 100049, Beijing, China}

\author[0000-0003-4856-2275]{Z. Chang}
\affiliation{Key Laboratory of Particle Astrophysics, Institute of High Energy Physics, Chinese Academy of Sciences, 100049, Beijing, China}

\author{J. Li}
\affiliation{CAS Key Laboratory for Research in Galaxies and Cosmology, Department of Astronomy, University of Science and Technology of China, Hefei 230026, China}
\affiliation{School of Astronomy and Space Science, University of Science and Technology of China, Hefei 230026, China}

\author{Wen T. Ye}
\affiliation{Key Laboratory of Particle Astrophysics, Institute of High Energy Physics, Chinese Academy of Sciences, 100049, Beijing, China}
\affiliation{University of Chinese Academy of Sciences, Chinese Academy of Sciences, 100049, Beijing, China}

\author{Pan P. Li}
\affiliation{Key Laboratory of Particle Astrophysics, Institute of High Energy Physics, Chinese Academy of Sciences, 100049, Beijing, China}
\affiliation{University of Chinese Academy of Sciences, Chinese Academy of Sciences, 100049, Beijing, China}

\author{Zhuo L. Yu}
\affiliation{Key Laboratory of Particle Astrophysics, Institute of High Energy Physics, Chinese Academy of Sciences, 100049, Beijing, China}

\author{Z. Yan}
\affiliation{Key Laboratory of Particle Astrophysics, Institute of High Energy Physics, Chinese Academy of Sciences, 100049, Beijing, China}
\affiliation{University of Chinese Academy of Sciences, Chinese Academy of Sciences, 100049, Beijing, China}
\affiliation{Yunnan Observatories, Chinese Academy of Sciences, Kunming 650216, P.R. China}
%% Note that the \and command from previous versions of AASTeX is now
%% depreciated in this version as it is no longer necessary. AASTeX 
%% automatically takes care of all commas and "and"s between authors names.

%% AASTeX 6.31 has the new \collaboration and \nocollaboration commands to
%% provide the collaboration status of a group of authors. These commands 
%% can be used either before or after the list of corresponding authors. The
%% argument for \collaboration is the collaboration identifier. Authors are
%% encouraged to surround collaboration identifiers with ()s. The 
%% \nocollaboration command takes no argument and exists to indicate that
%% the nearby authors are not part of surrounding collaborations.

%% Mark off the abstract in the ``abstract'' environment. 
\begin{abstract}
Although low frequency quasi-periodic oscillations (LFQPOs) are commonly detected in the X-ray light curves of accreting black hole X-ray binaries (BHXRBs), their origin still remains elusive. In this study, we conduct phase-resolved spectroscopy in a broad energy band for LFQPOs in MAXI J1820+070 during its 2018 outburst, utilizing Insight-HXMT observations. By employing the Hilbert-Huang Transform method, we extract the intrinsic QPO variability, and obtain the corresponding instantaneous amplitude, phase, and frequency functions for each data point. With well-defined phases, we construct QPO waveforms and phase-resolved spectra. By comparing the phase-folded waveform with that obtained from the Fourier method, we find that phase-folding on the phase of the QPO fundamental frequency leads to a slight reduction in the contribution of the harmonic component. This suggests that the phase difference between QPO harmonics exhibits time variability. Phase-resolved spectral analysis reveals strong concurrent modulations of the spectral index and flux across the bright hard state. The modulation of the spectral index could potentially be explained by both the corona and jet precession models, with the latter requiring efficient acceleration within the jet. Furthermore, significant modulations in the reflection fraction are detected exclusively during the later stages of the bright hard state. These findings provide support for the geometric origin of LFQPOs and offer valuable insights into the evolution of the accretion geometry during the outburst in MAXI J1820+070.
\end{abstract}

%% Keywords should appear after the \end{abstract} command. 
%% The AAS Journals now uses Unified Astronomy Thesaurus concepts:
%% https://astrothesaurus.org
%% You will be asked to selected these concepts during the submission process
%% but this old "keyword" functionality is maintained in case authors want
%% to include these concepts in their preprints.

%% From the front matter, we move on to the body of the paper.
%% Sections are demarcated by \section and \subsection, respectively.
%% Observe the use of the LaTeX \label
%% command after the \subsection to give a symbolic KEY to the
%% subsection for cross-referencing in a \ref command.
%% You can use LaTeX's \ref and \label commands to keep track of
%% cross-references to sections, equations, tables, and figures.
%% That way, if you change the order of any elements, LaTeX will
%% automatically renumber them.
%%
%% We recommend that authors also use the natbib \citep
%% and \citet commands to identify citations.  The citations are
%% tied to the reference list via symbolic KEYs. The KEY corresponds
%% to the KEY in the \bibitem in the reference list below. 

\section{Introduction} \label{sec:intro}
Black hole X-ray binaries (BHXRBs) are commonly observed as transient sources, with their primary feature being undergoing outbursts occasionally. Throughout an outburst, the X-ray emission properties of the system evolves significantly \citep{2006ARA&A..44...49R,2007A&ARv..15....1D}. Most complete outbursts typically exhibit four distinct canonical states: the low/hard state (LHS), hard intermediate state (HIMS), soft intermediate state (SIMS) and high soft state (HSS). Each state is characterized by distinct X-ray spectral and variability properties \citep[][]{2005A&A...440..207B,2005Ap&SS.300..107H}. In addition to X-ray emissions, BHXRBs are also characterized by radio/infrared emissions, which are generally believed to be associated with relativistic jets \citep[][and references therein]{2001MNRAS.322...31F,2004MNRAS.355.1105F,2022NatAs...6..577M}. In the hard state, the radio emission with a flat spectrum is interpreted as the self-absorbed synchrotron radiation from an optically thick, steady and compact jet \citep[][]{1979ApJ...232...34B,2001MNRAS.322...31F}. During the hard-to-soft transition, the compact jet is gradually quenched, and the radio emission is believed to originate from optically thin synchrotron radiation produced by transient plasma clouds with relativistic velocities \citep[][]{2004MNRAS.355.1105F}.

The X-ray spectrum of a BHXRB is typically composed of multiple components. First, there is a thermal component originating from a geometrically thin and optically thick accretion disc \citep{1973A&A....24..337S,1974MNRAS.168..603L}. Additionally, a non-thermal component arises from the Compton up-scattering of disc photons in a hot corona region ($\sim 100$ keV) \citep{1980A&A....86..121S,1994ApJ...434..570T,1996MNRAS.283..193Z,1999MNRAS.303L..11Z,1999MNRAS.309..561Z}, possibly extending to the jet base \citep{2005ApJ...635.1203M,2021NatCo..12.1025Y}. Furthermore, there is a reflection component resulting from a fraction of the Comptonized photons illuminating the disc and being scattered into the line of sight. The reflection spectrum exhibits prominent features such as broad Fe K$\alpha$ lines at $\sim6.4$ keV and a Compton hump in the $\sim$20--30 keV energy range, etc. \citep[][and references therein]{1980ApJ...236..928L,1989MNRAS.238..729F,2010MNRAS.409.1534D,2014ApJ...782...76G,2015ApJ...813...84G}.  

In the time domain, low frequency quasi-periodic oscillations \citep[LFQPOs, roughly 0.1--30 Hz,][]{1989ARA&A..27..517V} are usually seen in the light curves of BHXRBs, characterized by a narrow peak with finite width in the power density spectra (PDS). Based on the centroid frequency, quality factor and root-mean-square (rms) amplitude, LFQPOs are often classified into three types: A, B and C  \citep{1999ApJ...526L..33W,2005ApJ...629..403C,2006ARA&A..44...49R}. type-C QPOs are frequently observed in the LHS and HIMS, characterized by strong amplitudes (fractional rms $\sim10\%$) and the presence of flat-top noise components in the PDS.

Over the past few decades, several models have been proposed to explain the origin of LFQPOs in BHXRBs, based either on the geometric or the intrinsic properties of the accretion flow. \emph{Intrinsic} models include the trapped corrugation modes \citep{1990PASJ...42...99K,1999PhR...311..259W}, the Accretion-ejection instability model \citep[AEI,][]{1999A&A...349.1003T}, and the Two-Component Advection Flow model \citep{1996ApJ...457..805M}, etc.. \emph{Geometric} models, on the other hand, are mainly associated with the relativistic Lense-Thirring (L-T) precession, which was initially proposed as the dynamic mechanism for QPOs by \citet{1998ApJ...492L..59S}. As an extension of the relativistic precession model \citep[RPM;][]{1999ApJ...524L..63S}, the L-T precession model proposed by \citet{2009MNRAS.397L.101I} assumes the entire hot flow precesses within the inner radius of the truncated disc \citep{1997ApJ...489..865E}. Another \emph{geometric} model is the jet precession model \citep{2016MNRAS.460.2796S,2019MNRAS.485.3834D,2021NatAs...5...94M,2023ApJ...948..116M}, supported by the recent simulation \citep{2018MNRAS.474L..81L} and observations of synchronous QPO signals in X-ray and optical/IR bands \citep{2016MNRAS.460.3284K,2022MNRAS.513L..35T}.

Observational studies have revealed that the variability of type-C QPOs generally increases with photon energy \citep{2017ApJ...845..143Z,2018ApJ...866..122H,2020JHEAp..25...29K,2020MNRAS.494.1375Z} and no prominent disc-like component exists in the rms spectra \citep{2006MNRAS.370..405S,2014MNRAS.438..657A,2016MNRAS.458.1778A}. Additionally, \citet{2021NatAs...5...94M} reported the discovery of tpye-C QPOs above 200 keV, so the type-C QPO should be strongly related to the Comptonized emission. Moreover, the inclination dependence of amplitudes and time lags \citep[see][]{2015MNRAS.447.2059M,2017MNRAS.464.2643V} and reflection variability extracted from the phase-resolved spectroscopy \citep[see][]{2015MNRAS.446.3516I,2016MNRAS.461.1967I,2017MNRAS.464.2979I,2022MNRAS.511..255N} provide further support for a geometrical origin of LFQPOs, such as the L-T precession model. We refer readers to \citet{2019NewAR..8501524I} for recent reviews of observations and theories of LFQPOs.

Phase-resolved analysis provides valuable insights into the properties of QPOs by examining the phase-dependent behavior of the spectrum. In particular, the L-T precession model predicts that reflection features, such as emission lines, are modulated over the precession period \citep{2012MNRAS.427..934I,2016MNRAS.461.1967I,2020ApJ...897...27Y}. Therefore, conducting phase-resolved analysis of QPOs offers a powerful diagnostic tool to distinguish between different models. One possible method for generating phase-resolved spectra for QPOs is the Hilbert-Huang Transform \citep[HHT][]{1998RSPSA.454..903H}. The HHT is a powerful tool for analyzing phenomena with non-stationary periodicity and has been successfully applied in astronomical researches. For instance, it has been used to study the superorbital modulation of SMC X-1 \citep{2011ApJ...740...67H}, the 11-year sunspot variability \citep{2011SoPh..269..439B}, gravitational wave signals \citep{2007PhRvD..75f1101C,2022ApJ...935..127H}, as well as QPOs in the active galactic nucleus RE J1034+396 \citep{2014ApJ...788...31H}, and BHXRBs XTE J1550-564 \citep{2015ApJ...815...74S} and MAXI J1820+070 \citep{2023arXiv230512317Y}. The HHT enables us to not only trace the variation in frequency in the QPO but also to process phase-resolved analyses even though the periodicity is unstable. 

MAXI J1820+070 (ASASSN-18ey) is a galactic low-mass X-ray binary (LMXB) discovered in optical by All-Sky Automated Survey for Super Novae \citep[ASAS-SN, ][]{2014ApJ...788...48S} on 2018 March 3. Then it was discovered as an X-ray transient by MAXI/GCS on 2018 March 11 \citep{2018ATel11400....1D,2018ATel11399....1K,2018ApJ...867L...9T}, located at R.A.=$18^{\rm h}20^{\rm m}21^{\rm s}.9$ and decl.=$+07^\circ11'07''.3$ \citep{2018ATel11399....1K}. So far, it is one of the brightest X-ray transients, having a soft X-ray flux of $\sim4$ Crab in 2--4 keV \citep{2019ApJ...874..183S} with a low column density of $1.5\times10^{21} \rm cm^{-2}$ \citep{2018ATel11423....1U}. Based on the radio parallax measurement, \citet{2020MNRAS.493L..81A} determined a distance to the source of $2.96\pm0.33$ kpc. The VLBI observation provides an inclination angle of $63\pm3^\circ$ and a mass of $9.2\pm1.3 M_{\odot}$ \citep{2020MNRAS.493L..81A}. With more accurate measurement of the mass ratio, \citet{2020ApJ...893L..37T} gave a mass of $8.48^{+0.79}_{-0.72} M_{\odot}$. \citet{2021MNRAS.504.2168G} and \citet{2021ApJ...916..108Z} estimated the BH spin as $0.2^{+0.2}_{-0.3}$ and $0.14\pm0.09$, respectively, via modelling the continuum spectra. In this study, we investigate LFQPOs in MAXI J1820+070 in details by utilizing the HHT method to perform the phase-resolved spectral analysis. In Section~\ref{sec:2}, we describe the Insight-HXMT observations and data reductions. In Section~\ref{sec:3}, we firstly provide a concise introduction to the HHT method and its application in phase-resolved spectral analysis of QPOs, and then present the corresponding results. Subsequently, we discuss these results in Section~\ref{sec:4} and summarize in Section~\ref{sec:5}.

\section{Observations and Data Reduction} \label{sec:2}
In this study, we analyze data collected by the Hard X-ray Modulation Telescope (Insight-HXMT) \citep{2014SPIE.9144E..21Z,2020SCPMA..6349502Z}, which was launched on June 15, 2017, as the first Chinese X-ray astronomy satellite. The science payload of Insight-HXMT allows for observations across a broad energy band (1--250 keV) using three telescopes: the High Energy X-ray telescope (HE, composed of NaI/CsI, 20--250 keV), the Medium Energy X-ray telescope (ME, with a Si pin detector, 5--30 keV), and the Low Energy X-ray telescope (LE, using an SCD detector, 0.7--13 keV). These telescopes can operate in scanning and pointing observational modes as well as in GRB mode. For additional information about the Insight-HXMT mission, please refer to \citet{2019ApJ...879...61Z,2020SCPMA..6349503L,2020SCPMA..6349504C,2020SCPMA..6349505C}.

\begin{figure}
\centering
    \includegraphics[width=0.48\linewidth]{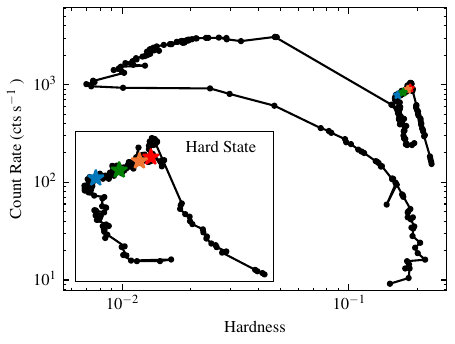}
    \caption{The hardness-intensity diagram (HID) of Insight-HXMT, where the hardness ratio is defined as the count ratio between the hard band (5--10 keV) and the soft band (1--5 keV), and the count rate is measured in the 1--10 keV energy range. The colored stars indicate the averaged results for each epoch.}
    \label{fig:1}
\end{figure}

Insight-HXMT detected numerous type-C QPOs in MAXI J1820+070 during the hard state of its 2018 outburst \citep[see][]{2021NatAs...5...94M,2022ApJ...932....7Y}. In this work, we select 30 observations that span over the bright hard state, in which the QPO frequency evolves from $\sim0.04$ to $\sim0.5$ Hz. In Figure~\ref{fig:1}, we plot the Hardness-intensity diagram (HID) for MAXI J1820+070 during its 2018 outburst. Based on the observational dates and evolution of the count rate, we categorize the selected 30 observations into four distinct epochs (plotted as colored stars in Figure~\ref{fig:1}). Specifically, Epochs 1, 2, 3 and 4 correspond to the time intervals MJD 58,204--MJD 58,209, MJD 58,220--MJD 58,225, MJD 58,228--MJD 58,239, and MJD 58,242--MJD 58,253, respectively. From Epochs 1 to 4, the count rate of LE evolves from $\sim 1000$ to $\sim 700$ cts s$^{-1}$. As a result, we combine more observations in Epochs 3 and 4 compared to Epochs 1 and 2, in order to achieve comparable count statistics within these four epochs for subsequent QPO phase-resolved spectral analysis. Detailed information regarding the aforementioned observations can be found in Table~\ref{tab:1}. We extract the event data using Insight-HXMT Data Analysis Software v2.05, along with the current calibration model v2.06\footnote{\url{http://hxmtweb.ihep.ac.cn/software.jhtml}} and the standard Insight-HXMT Data Reduction Guide v2.05\footnote{\url{http://hxmtweb.ihep.ac.cn/SoftDoc.jhtml}}, and apply the following series of criteria recommended by the Insight-HXMT team: (1) the elevation angle (ELV) is greater than $10^{\circ}$; (2) the geometric cutoff rigidity (COR) is greater than 8 GeV; (3) the pointing position offset is less than $0.04^{\circ}$; and (4) the good time intervals (GTIs) are at least 300 s away from the South Atlantic Anomaly (SAA). The backgrounds are produced from blind detectors using the LEBKGMAP, MEBKGMAP, and HEBKGMAP tools, version 2.0.9, based on the standard Insight-HXMT background models \citep{2020JHEAp..27...14L,2020JHEAp..27...24L,2020JHEAp..27...44G}. All photon are corrected to the barycenter of the solar system for their arrival times with the HXMTDAS tool \textit{hxbary}, and then evenly light curve with a bin size of 0.1 s for the subsequent timing and spectral analyses of QPOs. The spectral analysis is performed using the XSPEC v12.12.0 software package \citep{1996ASPC..101...17A}.

\begin{table}[]
    \centering
    \caption{Log of Insight-HXMT Observations Used in This Work\label{tab:1}}
    \begin{tabular}{cccccc}
    \hline
    \hline
    \# & ObsID & Start Time (MJD) & Exposure (s) & QPO Frequency (Hz) & Note \\
    \hline
    1 & P0114661006 & 58204.56 & 3325.7 & $0.046\pm0.001$ & Epoch 1 \\
    2 & P0114661007 & 58205.55 & 2550.6 & $0.048\pm0.003$ & Epoch 1 \\
    3 & P0114661008 & 58207.04 & 4649.3 & $0.051\pm0.002$ & Epoch 1 \\
    4 & P0114661009 & 58207.83 & 3728.7 & $0.050\pm0.003$ & Epoch 1 \\
    5 & P0114661010 & 58209.03 & 5356.6 & $0.055\pm0.002$ & Epoch 1 \\
    \hline
    6 & P0114661019 & 58220.82 & 9654.8 & $0.090\pm0.004$ & Epoch 2 \\
    7 & P0114661020 & 58222.61 & 7601.0 & $0.103\pm0.002$ & Epoch 2 \\
    8 & P0114661021 & 58223.20 & 2992.5 & $0.101\pm0.003$ & Epoch 2 \\
    9 & P0114661022 & 58224.48 & 2573.5 & $0.110\pm0.003$ & Epoch 2 \\
    10 & P0114661023 & 58225.48 & 4189.5 & $0.118\pm0.004$ & Epoch 2 \\
    \hline
    11 & P0114661026 & 58228.76 & 6492.8 & $0.135\pm0.004$ & Epoch 3 \\
    12 & P0114661030 & 58229.89 & 2137.6 & $0.142\pm0.005$ & Epoch 3 \\
    13 & P0114661031 & 58230.95 & 3247.9 & $0.155\pm0.004$ & Epoch 3 \\
    14 & P0114661027 & 58231.94 & 2813.0 & $0.168\pm0.005$ & Epoch 3 \\
    15 & P0114661028 & 58233.73 & 4639.4 & $0.173\pm0.005$ & Epoch 3 \\
    16 & P0114661029 & 58235.53 & 4215.4 & $0.199\pm0.006$ & Epoch 3 \\
    17 & P0114661032 & 58236.74 & 4029.9 & $0.259\pm0.011$ & Epoch 3 \\
    18 & P0114661033 & 58238.04 & 1436.5 & $0.254\pm0.007$ & Epoch 3 \\
    19 & P0114661034 & 58239.27 & 1460.4 & $0.278\pm0.007$ & Epoch 3 \\
    \hline
    20 & P0114661037 & 58242.22 & 1691.8 & $0.316\pm0.009$ & Epoch 4 \\
    21 & P0114661038 & 58243.60 & 2871.8 & $0.329\pm0.008$ & Epoch 4 \\
    22 & P0114661039 & 58244.61 & 1878.3 & $0.337\pm0.008$ & Epoch 4 \\
    23 & P0114661040 & 58245.61 & 2719.2 & $0.409\pm0.013$ & Epoch 4 \\
    24 & P0114661041 & 58246.60 & 5513.2 & $0.397\pm0.007$ & Epoch 4 \\
    25 & P0114661042 & 58247.73 & 2632.4 & $0.461\pm0.007$ & Epoch 4 \\
    26 & P0114661043 & 58248.72 & 2633.4 & $0.433\pm0.011$ & Epoch 4 \\
    27 & P0114661044 & 58250.24 & 3950.1 & $0.426\pm0.011$ & Epoch 4 \\
    28 & P0114661045 & 58251.03 & 2729.2 & $0.466\pm0.010$ & Epoch 4 \\
    29 & P0114661046 & 58252.56 & 2394.0 & $0.462\pm0.009$ & Epoch 4 \\
    30 & P0114661047 & 58253.15 & 1137.1 & $0.472\pm0.011$ & Epoch 4 \\
    \hline
    \hline
    \end{tabular}
\end{table}

\section{Analysis and Results}
\label{sec:3}
\subsection{Extract Intrinsic QPO Light Curve Using Variational Mode Decomposition} 
\begin{figure}
\centering
    \includegraphics[width=0.7\linewidth]{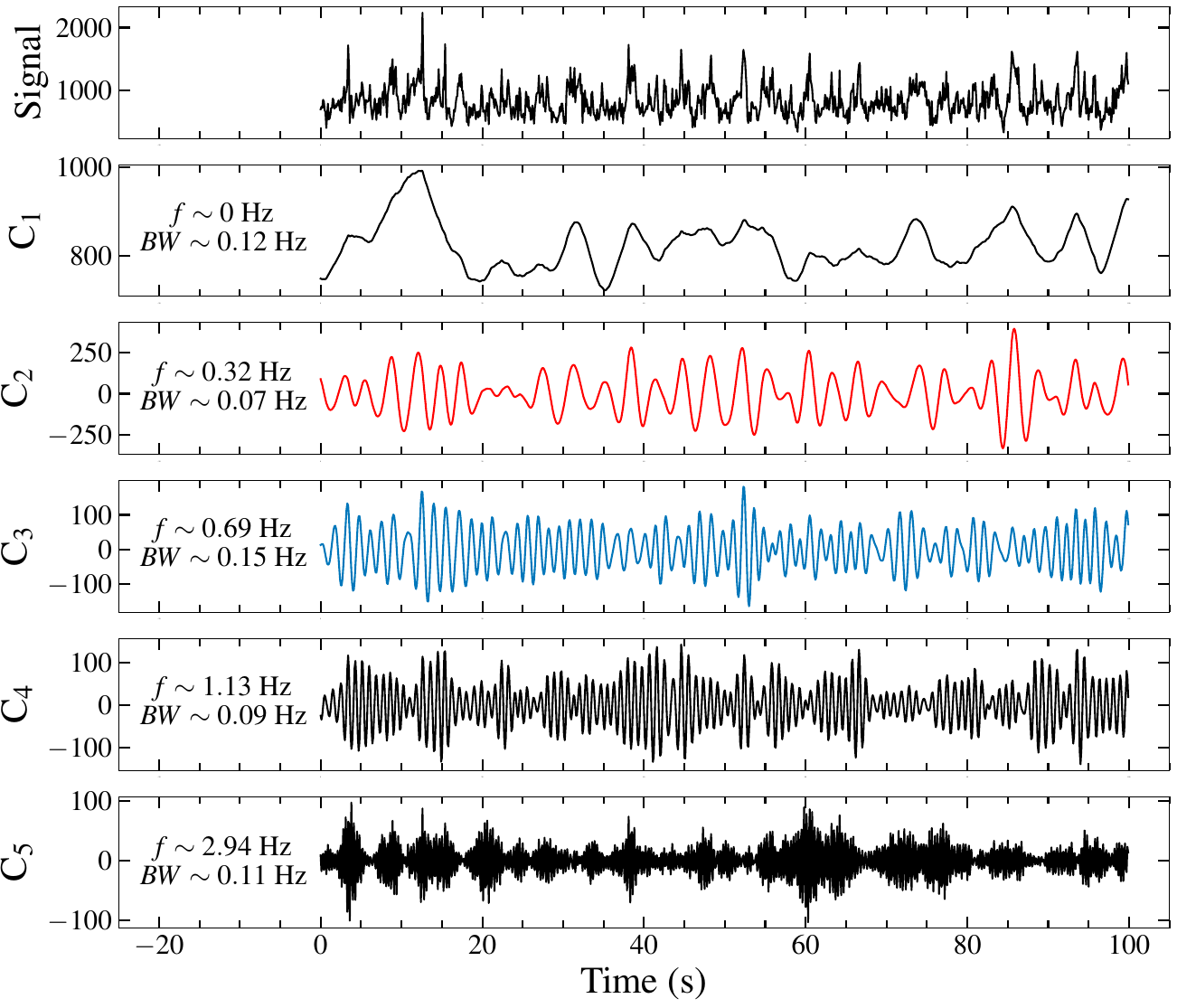}
    \caption{Representative example of a 100-s-long lightcurve of LE in energy range of 1--10 keV and its corresponding IMFs. In this example, we set $K = 5$ and $\alpha = 4000$ for the VMD algorithm, hence obtain five IMFs. The IMF C$_2$ corresponds to the QPO signal at $\sim0.3$ Hz, and the IMF C$_3$ corresponds to the second harmonic of the QPO. In this plot, the characteristic frequency ($f$) and the approximate bandwidth ($BW$) for each mode are also presented.}
    \label{fig:2}
\end{figure}

\begin{figure}
\centering
    \includegraphics[width=0.45\linewidth]{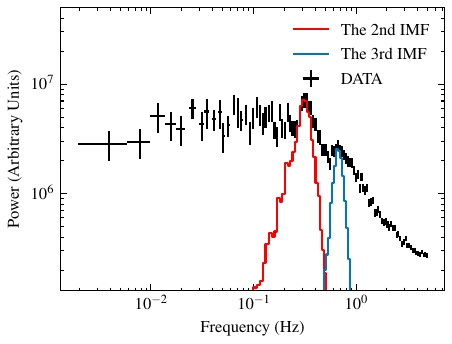}
    \caption{Power density spectra (PDS) of the original light curve (black dots), IMF C$_2$ (red line), and IMF C$_3$ (blue line) plotted in Figure~\ref{fig:2}.}
    \label{fig:3}
\end{figure}

Phase-resolved spectral analysis has been extensively performed in studies of eclipses and neutron star pulsations by folding the light curve. However, when dealing with QPO signals in black holes, simply folding the light curve is not appropriate due to the non-linear and non-deterministic evolution of their phase over time \citep[e.g.][]{1997ApJ...482..993M,2014ApJ...788...31H,2015MNRAS.446.3516I}. The Hilbert-Huang transform, an adaptive data analysis method developed by \citet{1998RSPSA.454..903H}, is a powerful tool to study signals with non-stationary periodicity \citep{1998RSPSA.454..903H,2008RvGeo..46.2006H}. This approach consists of two main components: mode decomposition and Hilbert spectral analysis (HSA). The mode decomposition aims to decomposing a time series into several intrinsic mode functions (IMFs), while the HSA allows obtaining both the frequency and phase functions of the desired IMFs, such as the QPO component \citep[e.g.][]{2014ApJ...788...31H,2020ApJ...900..116H}. The original method for mode decomposition is the empirical mode decomposition (EMD) proposed by \citep{1998RSPSA.454..903H}, which directly operates in the temporal space and adaptively decomposes the input signal into several IMFs. For a detailed algorithmic description of EMD, we recommend referring to the review by \citet{2008RvGeo..46.2006H}. The EMD is known to have limitations, such as the lack of mathematical theory and sensitivity to noise and sampling. In contrast, the variational mode decomposition (VMD) is a novel signal processing method that surpasses traditional decomposition approaches of e.g. EMD \citep{6655981}. Compared to the traditional EMD technique, VMD theoretically resolves the issue of mode mixing by decomposing the signal into a sum of IMFs with analytically calculated limited center frequency and bandwidth. In the VMD method, the IMFs are defined as amplitude-modulated-frequency-modulated (AM-FM) signals, expressed as:
\begin{equation}
u_k(t)=A_k(t)\cos{(\phi_k(t))},
\end{equation}
here the phase of the $k$-th mode, denoted as $\phi_k(t)$, is a non-decreasing function, indicating that the instantaneous frequency, $\omega_k(t)=\phi'_k(t)\geq0$. Additionally, the envelope $A_k(t)$ is non-negative ($A_k(t)\geq0$), while both the instantaneous frequency $\omega_k(t)$ and envelope $A_k(t)$ exhibit much slower variations compared to the phase $\phi_k(t)$ \citep{DAUBECHIES2011243,6522142}. \citet{6655981} demonstrated that a defined IMF is a signal with limited bandwidth, which serves as the main assumption for mode decomposition in the VMD. In other words, VMD assumes that each IMF is primarily concentrated around a center frequency $\omega_k$. For a mode, $u_k(t)$, with a mean frequency, $w_k$, its practical bandwidth increases both with the maximum deviation $\Delta f$ of the instantaneous frequency from its center, and with the rate of this excursion, $f_{\rm FM}$, according to Carson’s rule \citep{1666700}. In addition to this, the bandwidth of the envelope, $A_k(t)$, modulating the amplitude of the FM signal, itself given by its highest frequency, $f_{\rm AM}$, broadens the spectrum even further. Therefore, the total practical bandwidth of an IMF can be estimated as: $BW = 2(\Delta f + f_{\rm FM} + f_{\rm AM})$. To determine the bandwidth of each mode, VMD utilizes several signal processing tools. These include applying the Wiener filter for signal denoising, employing the Hilbert transform to construct a single-sideband analytic signal, and performing frequency shifting to baseband using complex harmonic mixing. We recommend to refer to \citet{6655981} for detailed information regarding these signal processing tools. The candidate mode is initially made single-sided, then shifted to baseband, and its bandwidth is estimated using $H^1$ Gaussian smoothness of the shifted signal, i.e. the $L^2$ norm of the gradient\footnote{In Equation (2), $||h(x)||_2$ means the $L^2$ (Euclidean) norm of the function $h(x)$, which is calculated as $\left[\int{|h(x)|^2dx}\right]^{1/2}$. For a discrete vector $h_k(x_k)$, the $L^2$ norm of it is calculated as $\left[\sum_{k=1}^K |h_k|^2\right]^{1/2}$. From another perspective, the inner product can be expressed as $\langle p(x), q(x)\rangle=\int{p^*(x)q(x)dt}$, which implies that $||h(x)||_2=\left[\langle h(x), h(x)\rangle\right]^{1/2}$.}:
\begin{equation}
BW_k = \left|\left|\partial_t\left[\left(\delta(t)+\frac{j}{\pi t}\right)*u_k(t)\right]e^{-j\omega_kt}\right|\right|_2,
\end{equation}
where $BW_k$ indicates the bandwidth of the mode, $\delta$ is the Dirac function, and `$*$' denotes convolution. The term enclosed in the square brackets represents the analytic signal of $u_k(t)$, which will be introduced in details in the subsequent text. For an input time series, $f(t)$, the VMD initializes the set of all modes as $\{u_k\}=\{u_1,...,u_K\}$ along with their corresponding  center frequencies $\{\omega_k\}=\{\omega_1,...,\omega_K\}$. The objective is to minimize the sum of bandwidths of the modes by solving the following optimization problem:
\begin{equation}
\min_{\{u_k\},\{\omega_k\}}\left\{\alpha\sum_{k=1}^K BW_k^2 + \left|\left|f(t)-\sum_{k=1}^K u_k(t)\right|\right|_2^2\right\},
\end{equation}
where $K$ is the total number of modes. The first term in the objective function promotes the compactness of the modes, while the second term measures the reconstruction error that needs to be minimized. The parameter $\alpha$ serves as a weighting factor that balances the compactness and reconstruction error. The VMD algorithm iteratively updates $\{u_k\}$ and $\{\omega_k\}$ until convergence. In this process, $\omega_k$ is directly calculated in the Fourier domain as:
\begin{equation}
    \omega_k = \frac{\int_0^\infty\omega \left|\hat{u}_k(\omega)\right|^2d\omega}{\int_0^\infty\left|\hat{u}_k(\omega)\right|^2d\omega},
\end{equation}
where $\hat{u}_k(\omega)$ represents the Fourier transform of the mode $u_k(t)$. It has been demonstrated that the solution of the optimization problem can be considered as a generalization of the Wiener filtering into adaptive, multiple bands. For more detailed description of the method, the complete constrained variational optimization problem is available in \citet{6655981}. The choice of $K$ and $\alpha$ is crucial in the VMD algorithm and determines its performance. Typically, higher values of $K$ are associated with higher values of $\alpha$, indicating that as the number of modes increases, the bandwidth of each mode decreases. In this study, our ultimate goal is to perform the phase-resolved spectral analysis for QPOs. Therefore, in the step of mode decomposition, our primary objective is to extract the intrinsic light curve of the QPO fundamental as one IMF. To accomplish this, we initially set $\alpha$ to a common value of 2000 and chose 2 modes ($K=2$). Subsequently, we calculate the power spectrum for each mode and compare it with the power spectrum of the original light curve. If none of the modes exhibit a PDS that matches the characteristic profile of the QPO, we increase the value of $K$. This iterative process continues until we successfully capture QPO's fundamental and harmonic modes in an individual mode. Finally, we adjust the value of $\alpha$ to match the profile of the PDS of the QPO IMF with that of the QPO peak in the PDS of the original lightcurve. %It is worth noting that slight adjustments to the initial values of $K$ and $\alpha$ do not significantly impact results of the subsequent phase-resolved analysis.
In this work, the code we used to perform VMD is from vmdpy v0.2\footnote{\url{https://github.com/vrcarva/vmdpy}} \citep{CARVALHO2020102073}, which is an open-source Python package. Figure~\ref{fig:2} shows an example of  a 100-s-long light curve from LE (obsID P0114661038) with its corresponding IMFs. The light curve exhibits a LFQPO with a frequency of $\sim$0.3 Hz. In this example, we set $K=5$ and $\alpha=4000$ for the VMD algorithm. A $\sim0.3$ Hz oscillation and its corresponding harmonic component are identified as the second (plotted in red) and third IMFs (plotted in blue), respectively. To provide a comparison with the Fourier analysis, we present power density spectra of the original LE light curve, as well as the second and third IMFs in the frequency range of 0.004 to 5 Hz in Figure~\ref{fig:3}.

\subsection{Hilbert Transform and QPO Waveform}
\begin{figure}
\centering
    \begin{minipage}[c]{0.46\textwidth}
\centering
    \includegraphics[width=\linewidth]{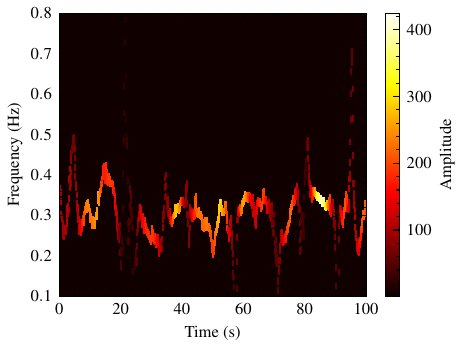}
\end{minipage}
\begin{minipage}[c]{0.46\textwidth}
\centering
    \includegraphics[width=\linewidth]{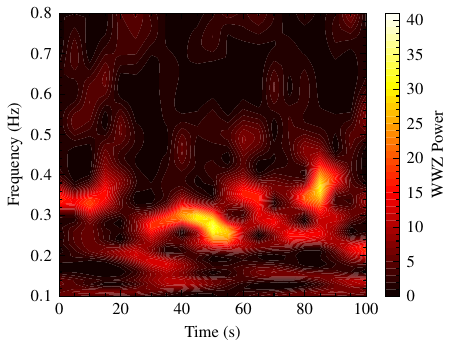}
\end{minipage}
    \caption{\textit{Left}: Hilbert spectrum of $\sim0.3$ Hz QPO IMF plotted in third panel of Figure~\ref{fig:2}. The color on the z-axis represents the amplitude. \textit{Right}: two-dimensional contour map of the WWZ power spectrum. Both of the two distinct methods present that the QPO frequency fluctuates between 0.2 and 0.4 Hz over time.}
    \label{fig:4}
\end{figure}

For each IMF, the Hilbert transform can provide physically meaningful phase, amplitude and frequency functions.
The Hilbert transform of a time series $f(t)$ is defined as 
\begin{equation}
\mathcal{H}(f(t))=\frac{1}{\pi}{\rm p.v.}\int{\frac{f(\tau)}{t-\tau}d\tau},
\end{equation}
where $\rm p.v.$ is the Cauchy principal value. Using the Hilbert transform, one can obtain the corresponding analytic signal, which is defined as 
\begin{equation}
    f_{\rm A}(t) = f(t) + j\mathcal{H}(f(t))=A(t)e^{j\phi(t)},
\end{equation}
where $j$ is the imaginary unit. The time-dependent amplitude, $A(t)$, and phase, $\phi(t)$, can be obtained from
\begin{equation}
A(t)=\left[f^2(t)+\mathcal{H}^2(f(t))\right]^{1/2}    
\end{equation}
and
\begin{equation}\label{eq:phi}
\phi(t)=\arctan{\left[\frac{\mathcal{H}(f(t))}{f(t)}\right]},
\end{equation}
respectively. Therefore, the instantaneous frequency, $\nu(t)$, can be defined as
\begin{equation}
\nu(t)=\frac{1}{2\pi}\frac{d\phi(t)}{dt}.
\end{equation}
Considering the focus of the present study on investigating QPOs, we solely apply the Hilbert spectral analysis to the QPO IMF (e.g. the second IMF plotted in Figure~\ref{fig:2}). The left panel of Figure~\ref{fig:4} illustrates the Hilbert spectrum, represented by a color map showcasing the instantaneous frequency and amplitude of the QPO component. The color depth represents the amplitude. For comparison, we also employ the Weight Wavelet Z-transform \citep[WWZ,][]{1996AJ....112.1709F} technique for the time-frequency analysis. The right panel of Figure~\ref{fig:4} displays the color contour of the WWZ power spectrum. In the WWZ analysis, we adopt specific parameters, including a restricted frequency range of 0.08--1 Hz, a frequency step of 0.01 Hz, and a decay constant of $c = 0.005$ \citep[details on these parameter settings can be found in ][]{1996AJ....112.1709F,2023ApJ...943..157L}. By comparing the Hilbert spectrum with the WWZ power map, it is evident that the QPO frequency fluctuates between 0.2 and 0.4 Hz over time. Importantly, the main structures obtained from these two distinct methods are in agreement. Both two methods show that, the $\sim$0.3 Hz oscillation is relatively weak in time intervals of $\sim$ 18--25 s and $\sim$ 65--80 s. However, it becomes stronger during time intervals of $\sim$ 40--60 s and $\sim$ 80--90 s, with frequencies of $\sim0.27$ and $\sim0.35$ Hz, respectively (see Figure~\ref{fig:4}).  

\begin{figure}
\centering
    \includegraphics[width=0.45\linewidth]{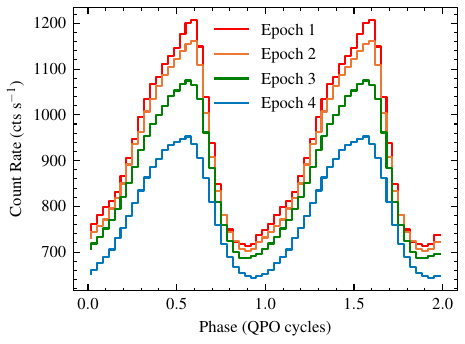}
    \caption{Constructed QPO waveforms of four epochs by phase-folding the lightcurve from LE in the energy range of 1--10 keV.} \label{fig:5}
\end{figure}

Once the instantaneous phases, $\phi(t)$, of the QPOs have been computed using the Hilbert transform as described in Equation (\ref{eq:phi}), the QPO waveforms can be constructed by phase-folding the LE light curves \citep[also see][]{2014ApJ...788...31H,2020ApJ...900..116H}. To generate the QPO waveform, we divide the phases into 30 bins. Figure~\ref{fig:5} illustrates the results for the four epochs. The QPO waveform exhibits a distinctive non-sinusoidal nature, characterized by a gradual rise of $\sim$0.7 cycles, followed by a rapid drop of $\sim$0.3 cycles. This non-sinusoidal behavior is associated with the presence of the second harmonic in the PDS (see Figure~\ref{fig:3}).

\subsection{Phase-resolved Spectra}
With the well-defined phase, we construct Good Time Intervals (GTIs) corresponding to 10 distinct phase intervals. These GTIs enable us to extract spectra for subsequent phase-resolved spectral analysis in the four epochs. For the spectral analysis, energy bands of 2--10 keV for LE, 10--30 keV for ME, and 30--120 keV for HE, respectively, are adopted. Due to calibration uncertainties at 22 keV in the ME, we exclude the 20--23 keV range during spectral analysis for this instrument. To account for the current accuracy of the instrument calibration, we introduce systematic errors of 0.5\%, 0.5\%, and 1\% to the energy spectra for LE, ME, and HE, respectively. We employ the same spectral model as in \citet{2021NatCo..12.1025Y}, which is $constant\times Tbabs\times(diskbb+relxillCp+xillverCp)$, to fit these spectra at different phases. The constant factor is utilized to accommodate flux calibration discrepancies among the three distinct instruments. In phase-averaged spectral fittings, we set the constant factor of LE to 1 and treat those of ME and HE as free parameters, allowing them to vary during the fittings. The $Tbabs$ accounts for interstellar absorption, with the equivalent hydrogen column, $N_{\rm H}$, fixed at $0.15\times10^{22}$ cm$^{-2}$ \citep{2018ATel11423....1U}. In the non-relativistic multi-color blackbody model $diskbb$, two crucial parameters are estimated, namely the temperature ($T_{\rm in}$) and the inner radius ($R_{\rm in}$) of the accretion disk. The latter one is linked to the normalization of the $diskbb$ model follows as $N_{\rm disk} = (R_{\rm in}/D_{10})^2\cos i$, where $N_{\rm disk}$ is the normalization, $R_{\rm in}$ is in units of km, $D_{10}$ is the distance to the source in units of 10 kpc and $i$ is the inclination. By taking the black hole mass as $8.48\ M_{\odot}$ \citep{2020ApJ...893L..37T}, the distance and inclination as 2.96 kpc and 63$^\circ$ , respectively \citep{2020MNRAS.493L..81A}, we can obtain $N_{\rm disk} \approx 816(R_{\rm in}/R_{\rm g})^2$, where $R_{\rm g}=GM/c^2$ is the gravitational radius. For the relativistic reflection model $relxillCp$, we assume the canonical power-law emissivity profile, $\epsilon=r^{-3}$, for the disk \citep{1989MNRAS.238..729F}. Referring to previous studies, the disk inclination ($i$), spin parameter of the black hole ($a*$) are set to $63^{\circ}$, 0.14, respectively \citep{2020MNRAS.493L..81A,2021ApJ...916..108Z}. The parameter $R_{\rm out}$ is the outer radius of the accretion disk, which turns out to not be sensitive to the overall fitting and hence is frozen at the maximum value, 1000$R_{\rm g}$. Since $R_{\rm in}$ can also be a free parameter in $relxillCp$, it is self-consistent to link it to $N_{\rm disk}$ in $diskbb$ as $N_{\rm disk}=816R_{\rm in}^2$ \citep[see also][]{2023MNRAS.520.5544G}, where $R_{\rm in}$ is in units of $R_{\rm g}$. However, it is important to emphasize that, for the sake of simplicity, we have disregarded any relativistic effects when establishing the connection between the two parameters. The electron density ($N_{\rm e}$) is set to the default value of 10$^{15}$ cm$^{-3}$ since our energy coverage (2--120 keV) does not allow for the constraint of soft excess emission below 2 keV \citep{2016MNRAS.462..751G}.
In the spectral fitting, we also apply the non-relativistic reflection model $xillverCp$ to account for the distant reflector with the narrow/low-ionization iron line \citep[e.g.,][]{2015ApJ...813...84G}. The parameters in $xillverCp$ are tied to the corresponding ones in $relxillCp$ except for the ionization parameter and the normalization. Since the spectral fits are insensitive to the ionization parameter, we fix the ionization parameter at a low value with $\log\xi=1.0$ \citep[see also][]{2021NatCo..12.1025Y}. Hence, the normalization of $xillverCp$ is the only free parameter in the model. 
In our phase-resolved spectral fittings, we fix constant values, the iron abundance $A_{\rm Fe}$ and the ionization parameter $\log\xi$ to the same values obtained from the phase-averaged spectral fittings. After fitting the energy spectra with the model, we obtain a satisfactory reduced $\chi^2$ ($\sim 1$). To compute the uncertainties at the 90\% confidence level, we employ the Markov Chain Monte Carlo (MCMC) method using the Goodman-Weare algorithm with 8 walkers and a total length of 40,000 \citep{2010CAMCS...5...65G}, and the initial 2000 elements are discarded as “burn-in” period. We find that the autocorrelation length is typically one thousand elements, so the net number of independent samples in the parameter space we have is order of $10^4$. In order to further test the convergence of the MCMC chain, we compare the one- and two-dimensional projections of the posterior distributions for each parameters from the first and second halves of the chain \citep[see also][]{2021ApJ...909...63L}, and we find no significant differences. In Appendix~\ref{appenix}, we present the contour maps and probability distributions for each free parameter.
%The best-fit parameters for the trough (0--0.1 cycles) and peak (0.5--0.6 cycles) phases in the four epochs are presented in Table 2.

\begin{figure}
\centering
\begin{minipage}[c]{0.212\textwidth}
\centering
    \includegraphics[width=\linewidth]{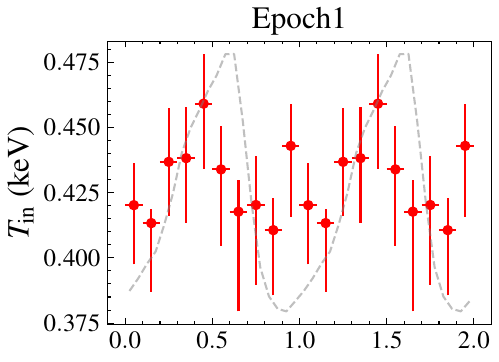}
\end{minipage}
\begin{minipage}[c]{0.2\textwidth}
\centering
    \includegraphics[width=\linewidth]{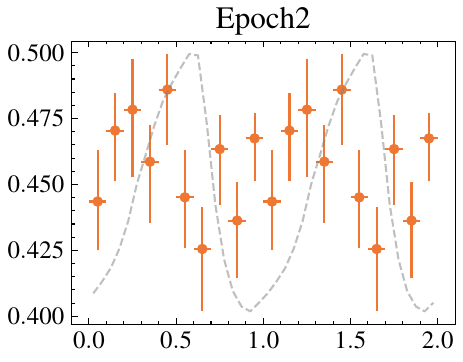}
\end{minipage}
\begin{minipage}[c]{0.2\textwidth}
\centering
    \includegraphics[width=\linewidth]{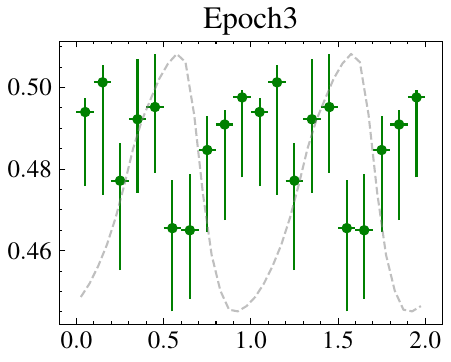}
\end{minipage}
\begin{minipage}[c]{0.2\textwidth}
\centering
    \includegraphics[width=\linewidth]{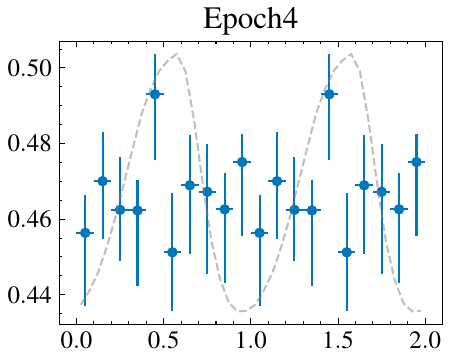}
\end{minipage}\\
\begin{minipage}[c]{0.212\textwidth}
\centering
    \includegraphics[width=\linewidth]{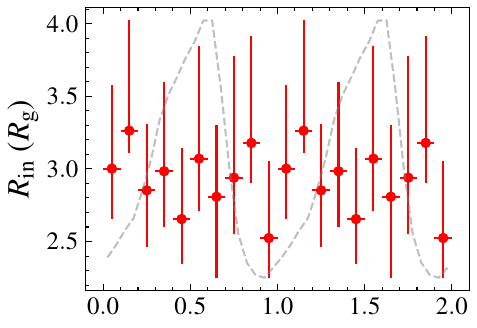}
\end{minipage}
\begin{minipage}[c]{0.2\textwidth}
\centering
    \includegraphics[width=\linewidth]{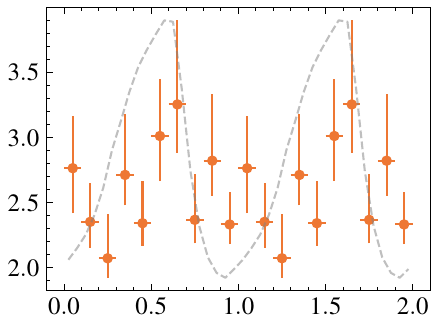}
\end{minipage}
\begin{minipage}[c]{0.2\textwidth}
\centering
    \includegraphics[width=\linewidth]{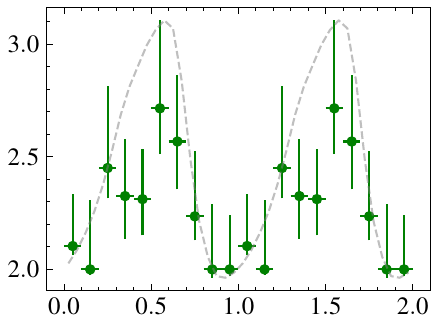}
\end{minipage}
\begin{minipage}[c]{0.2\textwidth}
\centering
    \includegraphics[width=\linewidth]{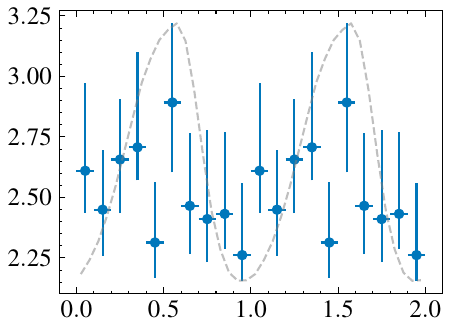}
\end{minipage}\\
\begin{minipage}[c]{0.212\textwidth}
\centering
    \includegraphics[width=\linewidth]{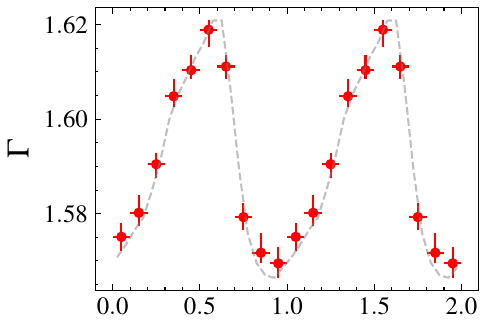}
\end{minipage}
\begin{minipage}[c]{0.2\textwidth}
\centering
    \includegraphics[width=\linewidth]{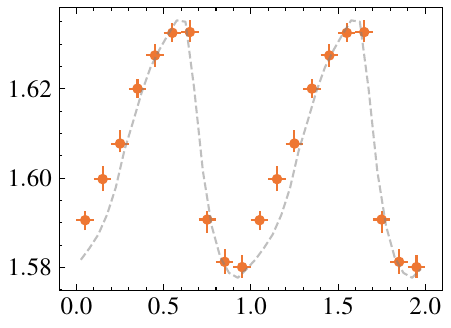}
\end{minipage}
\begin{minipage}[c]{0.2\textwidth}
\centering
    \includegraphics[width=\linewidth]{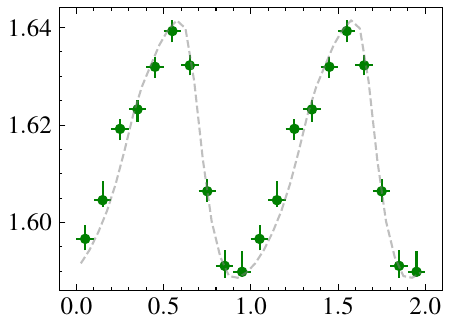}
\end{minipage}
\begin{minipage}[c]{0.2\textwidth}
\centering
    \includegraphics[width=\linewidth]{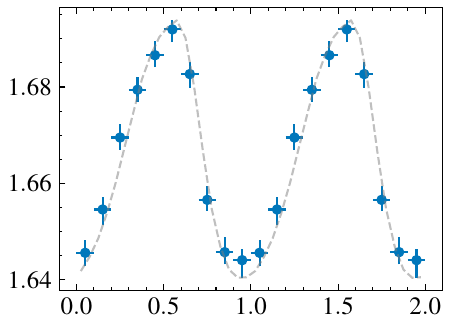}
\end{minipage}\\
\begin{minipage}[c]{0.212\textwidth}
\centering
    \includegraphics[width=\linewidth]{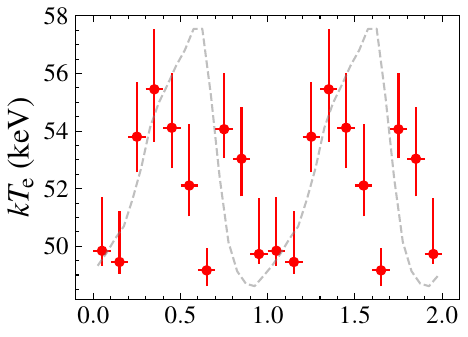}
\end{minipage}
\begin{minipage}[c]{0.2\textwidth}
\centering
    \includegraphics[width=\linewidth]{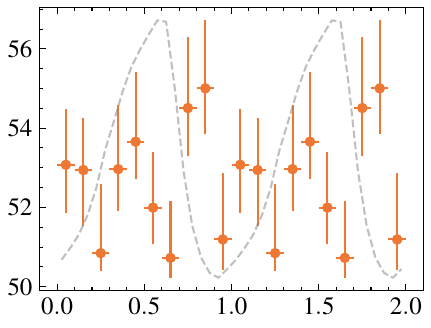}
\end{minipage}
\begin{minipage}[c]{0.2\textwidth}
\centering
    \includegraphics[width=\linewidth]{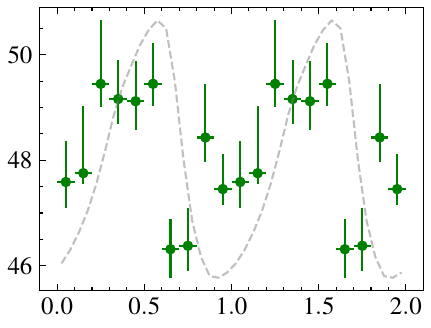}
\end{minipage}
\begin{minipage}[c]{0.2\textwidth}
\centering
    \includegraphics[width=\linewidth]{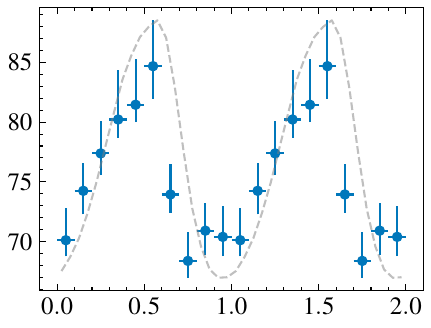}
\end{minipage}\\
\begin{minipage}[c]{0.212\textwidth}
\centering
    \includegraphics[width=\linewidth]{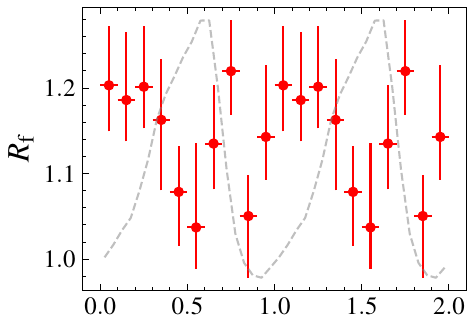}
\end{minipage}
\begin{minipage}[c]{0.2\textwidth}
\centering
    \includegraphics[width=\linewidth]{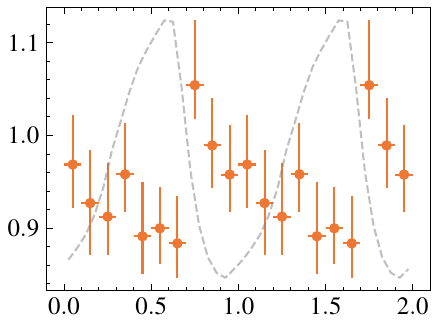}
\end{minipage}
\begin{minipage}[c]{0.2\textwidth}
\centering
    \includegraphics[width=\linewidth]{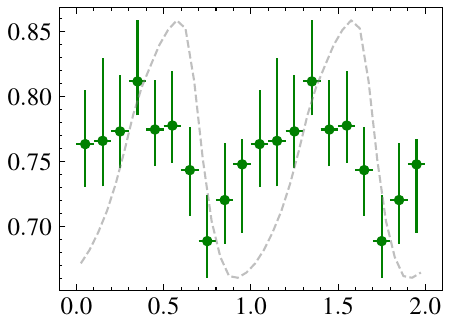}
\end{minipage}
\begin{minipage}[c]{0.2\textwidth}
\centering
    \includegraphics[width=\linewidth]{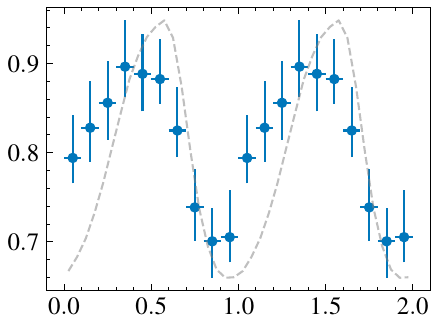}
\end{minipage}\\
\begin{minipage}[c]{0.212\textwidth}
\centering
    \includegraphics[width=\linewidth]{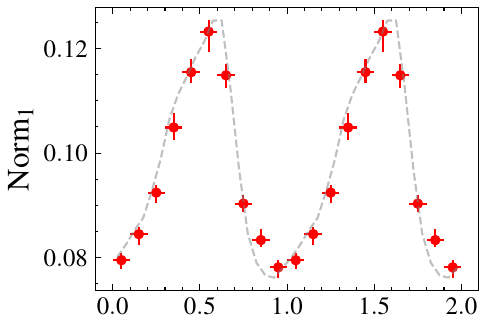}
\end{minipage}
\begin{minipage}[c]{0.2\textwidth}
\centering
    \includegraphics[width=\linewidth]{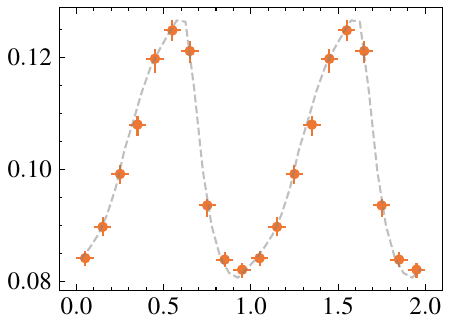}
\end{minipage}
\begin{minipage}[c]{0.2\textwidth}
\centering
    \includegraphics[width=\linewidth]{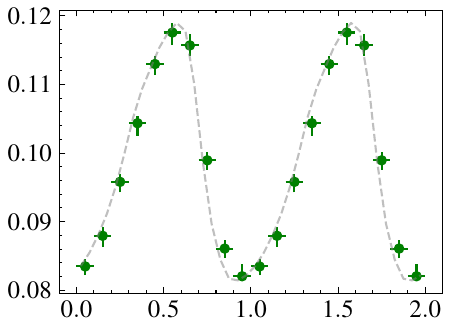}
\end{minipage}
\begin{minipage}[c]{0.2\textwidth}
\centering
    \includegraphics[width=\linewidth]{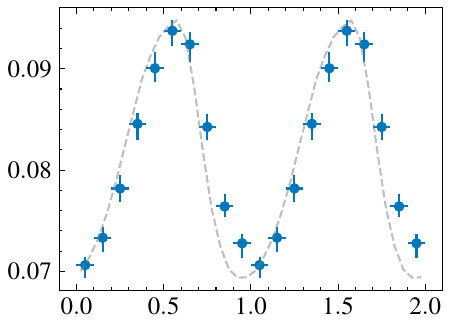}
\end{minipage}\\
\begin{minipage}[c]{0.212\textwidth}
\centering
    \includegraphics[width=\linewidth]{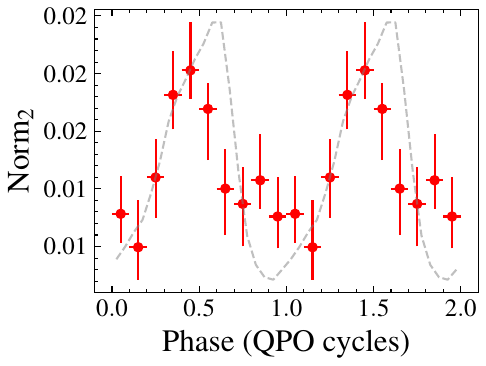}
\end{minipage}
\begin{minipage}[c]{0.2\textwidth}
\centering
    \includegraphics[width=\linewidth]{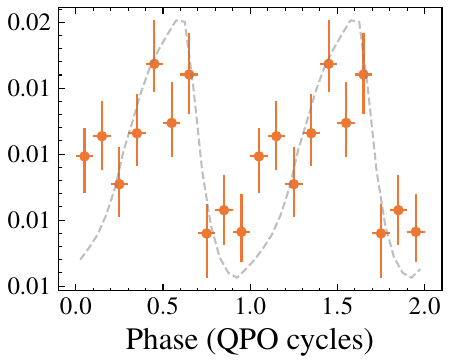}
\end{minipage}
\begin{minipage}[c]{0.2\textwidth}
\centering
    \includegraphics[width=\linewidth]{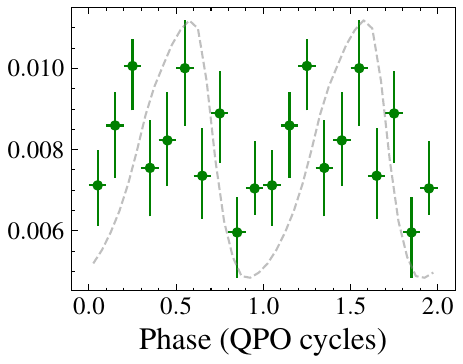}
\end{minipage}
\begin{minipage}[c]{0.2\textwidth}
\centering
    \includegraphics[width=\linewidth]{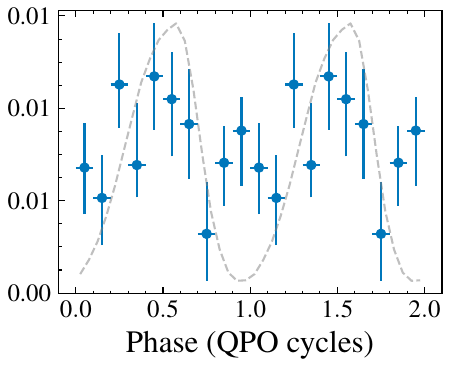}
\end{minipage}
    \caption{Phase dependence of the seven free spectral parameters, including the temperature of the inner radius of the disk ($T_{\rm in}$), inner radius ($R_{\rm in}$), spectral index ($\Gamma$), electron temperature ($kT_{\rm e}$), reflection fraction ($R_{\rm f}$), and the normalization of $relxillCp$ (Norm$_1$) and $xillverCp$ (Norm$_2$), across the four epochs. Additionally, we include the QPO waveform in each panel as the gray dashed line.}
    \label{fig:6}
\end{figure}

Figure~\ref{fig:6} displays the phase dependence of the seven free parameters, including the temperature of the inner radius of the disk ($T_{\rm in}$), inner radius ($R_{\rm in}$), spectral index ($\Gamma$), electron temperature ($kT_{\rm e}$), reflection fraction ($R_{\rm f}$), and the normalization of $relxillCp$ (Norm$1$) and $xillverCp$ (Norm$2$), across the four epochs. Additionally, we include the QPO waveform in each panel as the gray dashed line. From the phase-resolved spectral analysis, we observe no significant modulations in $T_{\rm in}$ and $R_{\rm in}$ as the flux varies on the QPO period. However, notable modulations are observed in the spectral index and the normalization of $relxillCp$ throughout all four epochs. Moreover, these two parameters exhibit the same modulation phase as the flux. Additionally, significant phase modulations are observed for parameters $R_{\rm f}$ and $kT_{\rm e}$ in Epochs 3 and 4, but not 1 and 2. For the modulation of $R_{\rm f}$ in Epochs 3 and 4, it seems to lead the flux by $\sim$0.2 cycles. On the other hand, the normalization of $xillverCp$ evolves from strong phase modulation in Epochs 1 and 2 to relatively marginal modulation in Epochs 3 and 4.

\section{Discussion} \label{sec:4}
In this work, we have analyzed observations of MAXI J1820+070 from Insight-HXMT and extracted the intrinsic QPO variability by utilizing the VMD algorithm to the observed light curves. Through the application of the Hilbert transform to the intrinsic QPO variability, the instantaneous amplitude, phase and frequency functions are obtained for each data point. The HSA reveals that the QPO exhibits variations in its frequency and amplitude. Via phase-revolved spectral analyses we find strong and persistent phase modulation of the spectral index, and moderate evolution of phase modulation in the reflection fraction. These findings provide valuable insights into the origins of LFQPOs and the evolution of accretion geometry during outbursts in BHXRB systems.

\subsection{QPO Waveform}
\begin{figure}
\centering
\begin{minipage}[c]{0.4\textwidth}
\centering
    \includegraphics[width=\linewidth]{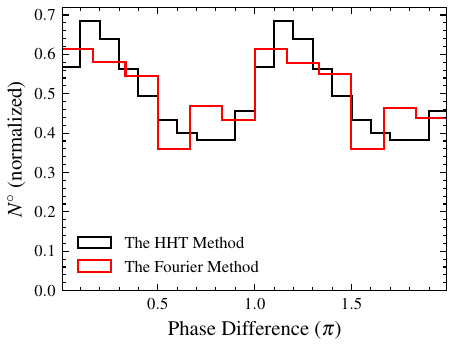}
\end{minipage}
\begin{minipage}[c]{0.4\textwidth}
\centering
    \includegraphics[width=\linewidth]{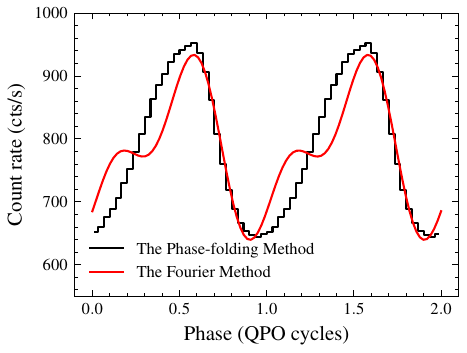}
\end{minipage}
\caption{\textit{Left}: Histogram of measured values of the phase difference between harmonics obtained using the HHT (black) and Fourier (red) methods. To show a clear peak, we repeat the data from 0 to $\pi$. \textit{Right}: Reconstructed QPO waveform derived the HHT phase-folding (black) and Fourier (red) methods.}
    \label{fig:waveform}
\end{figure}

In Section~\ref{sec:3}, the QPO waveforms are constructed through the phase-folding of light curves (see Figure~\ref{fig:5}). Despite that the non-sinusoidal nature, characterized by a distinct ``slow rise and fast decay" feature, is observed in the QPO waveforms across the four epochs, the contribution of the harmonic component in these waveforms is relatively weaker than that observed in PDS. The phase difference between QPO fundamental and harmonic modes has been demonstrated to be unstable in time \citep{2015MNRAS.446.3516I,2019MNRAS.485.3834D}. This indicates that the phase-folding on the phase of the fundamental mode could suppress the contribution of the harmonic component in the derived waveform. 
To demonstrate this suppressing effect in the phase-folding method, we compare the phase-folded waveform with that obtain from the Fourier method. To do this, we conduct a similar analysis as \citet{2015MNRAS.446.3516I} to reconstruct the QPO waveform for Epoch 4. In order to measure the phase of the QPO fundamental and harmonic components, we first perform the Fourier transform of many short segments of each light curve within Epoch 4. The phases of the fundamental and harmonic components for a single segment are calculated as the argument of the Fourier transform for that segment at the frequency bins containing the QPO's fundamental and harmonic components, respectively. Specifically, for the harmonic component, we utilize the frequency bin that is precisely twice the frequency of the bin used for the fundamental component. The phase difference for each segment is calculated using Equation (3) of \citet{2015MNRAS.446.3516I}. The distribution of the phase difference ($\psi/\pi$) is presented in the left panel of Figure~\ref{fig:waveform}, where the results obtained from the Fourier method is plotted in red. 
On the other hand, the HHT method enables us to obtain the instantaneous phase for each IMF. Thus, we calculate the phase for the IMFs that corresponds to the QPO's fundamental and harmonic components, respectively, and then determine the phase difference for each time bin ($dt=0.1$ s). The resulted $\psi/\pi$ distribution derived using the HHT method is plotted in black in the left panel of Figure~\ref{fig:waveform}. As one can see, both methods yield similar $\psi/\pi$ distributions, but the HHT method produces a more concentrated distribution. The mean values of $\psi/\pi$ from the Fourier and HHT methods are found to be $0.191\pm0.041$ and $0.229\pm0.002$, respectively. We propose that the HHT method could serve as an effective and novel approach to determine the phase difference between QPO harmonics. Based on the RMS of the fundamental and harmonic components, $\psi/\pi$, as well as the Equation (6) of \citet{2015MNRAS.446.3516I}, the QPO waveform can be reconstructed. The waveforms obtained from both the phase-folding and Fourier methods are presented in the right panel of Figure~\ref{fig:waveform}. Although both waveforms exhibit a non-sinusoidal nature, characterized by a ``slow rise and fast decay" feature, the contribution of the harmonic component in the waveform derived from the HHT method is indeed suppressed. It is worth noting that this suppressing effect may also introduce slight differences in the modulations of spectral parameters between the HHT and Fourier methods. 

\citet{2019MNRAS.485.3834D} measured the phase difference for many observations of QPOs in the RXTE archive. In most high-inclination systems, $\psi/\pi$ between harmonics is distributed around $\sim0$ to $\sim0.4$. This suggests that the majority of QPO waveforms observed in these systems exhibit the characteristic of ``slow rise and fast decay", even though the evolution of $\psi$ may slightly alter the waveform. Considering that the inclination of MAXI J1820+070 is $\sim63^\circ$, the QPO waveforms observed in these four epochs, which display the ``slow rise and fast decay" feature, are consistent with the behavior observed in most high-inclination systems from RXTE observations. In Epoch 4, we observe $\psi/\pi\sim0.2$ for QPOs at $\sim 0.4$ Hz, which is slightly smaller than the values observed in other high-inclination systems at similar frequencies. However, in order to make a comprehensive comparison of the evolution of $\psi$ in MAXI J1820+070 with that in other sources, it is necessary to have more QPO samples at higher QPO frequencies ($\sim1-5$ Hz) which were not detected by Insight-HXMT.

\subsection{Modulation of the Spectrum}
As shown in Figure~\ref{fig:6}, the spectral index ($\Gamma$) and the normalization of $relxillCp$ (Norm$_1$) exhibit significant modulations, which closely follow the flux modulation. In contrast, the modulations of parameters in $diskbb$ are marginal. This suggests that the dominant contribution to the QPOs comes from the Comptonized component, supporting models such as the corona's L-T precession model \citep{2009MNRAS.397L.101I} and the time-dependent Comptonization model \citep{2020MNRAS.492.1399K,2022MNRAS.515.2099B}. Since a jet could also act like a corona which produces Comptonized photons \citep{2005ApJ...635.1203M}, the jet precession model is also a plausible explanation for the findings \citep{2016MNRAS.460.2796S,2018MNRAS.474L..81L,2021NatAs...5...94M,2023ApJ...948..116M}. 

\subsubsection{Corona Precession Model}
If the QPOs are produced by the corona's Lense-Thirring precession, the X-ray QPO variability is due to the geometric wobble of the corona, which changes the projection area of the corona with respect to the observer to modulate the observed X-ray flux \citep{2009MNRAS.397L.101I,2015ApJ...807...53I,2018ApJ...858...82Y,2020ApJ...897...27Y,2023ApJ...943..165S}.

%\subsubsection{Modulation of Spectral Index}

\begin{figure}
\centering
\begin{minipage}[c]{0.24\textwidth}
\centering
    \includegraphics[width=\linewidth]{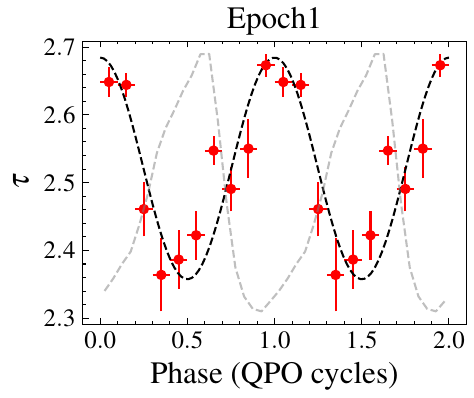}
\end{minipage}
\begin{minipage}[c]{0.24\textwidth}
\centering
    \includegraphics[width=\linewidth]{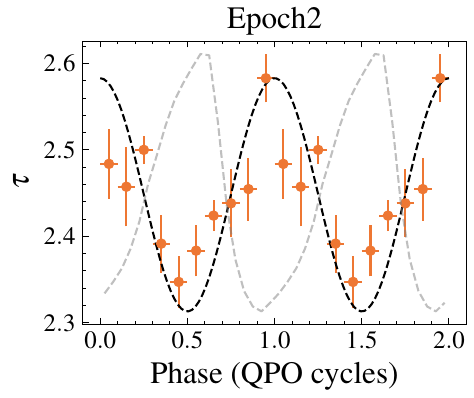}
\end{minipage}
\begin{minipage}[c]{0.24\textwidth}
\centering
    \includegraphics[width=\linewidth]{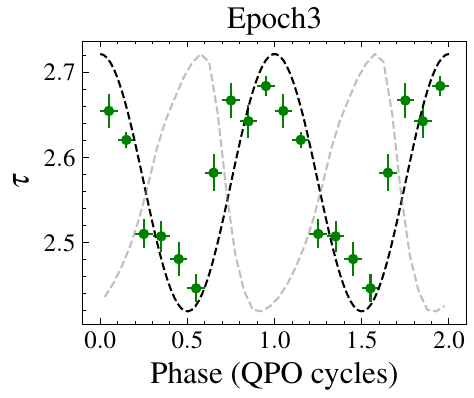}
\end{minipage}
\begin{minipage}[c]{0.24\textwidth}
\centering
    \includegraphics[width=\linewidth]{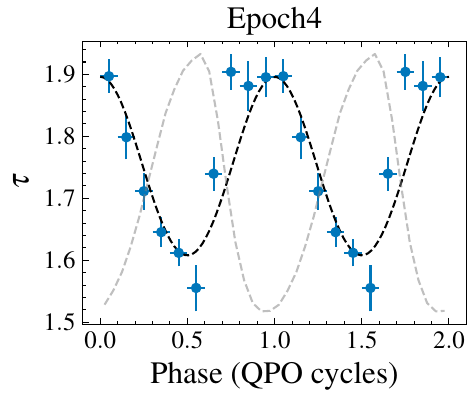}
\end{minipage}
\caption{Phase dependence of the optical depth across the four epochs. The colored scatters represent the calculated optical depth ($\tau$) from Equation~\ref{eq:tau}, using the best-fit parameters of $\Gamma$ and $kT_{\rm e}$ from the spectral fitting. The black dashed lines represent the calculated optical depth based on Equation~\ref{eq:9} and~\ref{eq:theta}.}
    \label{fig:7}
\end{figure}
Since the geometric shape of a corona might not be a sphere \citep[e.g., a crushed sphere,][]{2012MNRAS.427..934I,2023ApJ...943..165S}, the optical depth of the corona respect to the observer could vary over the precession period. The illuminating continuum in $relxillCp$ is $nthComp$ \citep{1999MNRAS.309..561Z}, wherein the spectral shape is determined by the combination of the electron temperature ($kT_{\rm e}$) and the optical depth ($\tau$). Therefore, the modulations of $\Gamma$ could be attributed to the modulations of $kT_{\rm e}$ and $\tau$. For a given $kT_{\rm e}$ and spectral index ($\Gamma$), we can estimate the corona's optical depth by 
\begin{equation}\label{eq:tau}
    \tau = \sqrt{\frac{9}{4}+\frac{m_{\rm e}c^2}{kT_{\rm e}}\frac{3}{(\Gamma-1)(\Gamma+2)}}-\frac{3}{2},
\end{equation}
where $m_{\rm e}$ is the electron rest mass, $c$ is the speed of light \citep{1996MNRAS.283..193Z}. We utilize Equation~(\ref{eq:tau}) to calculate the optical depth at different phases, using the best-fit parameters of $\Gamma$ and $kT_{\rm e}$ from the spectral fittings. The modulations of the optical depth for four epochs are plotted in Figure~\ref{fig:7} as colored scatters. For all four epochs, the optical depth exhibits contrasting modulations with respect to the flux. This phenomenon can be qualitatively explained by the corona's Lense-Thirring precession model. When the corona is oriented face-on towards the observer, the projected area and consequently the flux of the corona appear to be larger compared to the edge-on orientation. However, the optical depth of the corona modulates in the opposite manner due to the self-occultation effect \citep{2018ApJ...858...82Y}.  
Quantitatively, according to simplified calculations by \citet{2023ApJ...943..165S}, if the corona's shape resembles a crushed sphere, there might exist an anti-correlation between the flux and corona's optical depth. Equation (9) of \citet{2023ApJ...943..165S} describes the modulation of the corona's optical depth as the corona precesses:
\begin{equation}
\label{eq:9}
\tau=\tau_0\left\{\left(h/r\right)^2+\left[1-\left(h/r\right)^2\right]\cos^2{\theta}\right\}^{-1/2},
\end{equation}
where $\tau_0$ is the minimum optical depth of the corona, i.e. viewed from the corona's normal, and $h/r$ is the scale height which describes the shape of the corona. When $h/r=1$, the corona's shape is a sphere. $\theta$ is the included angle between the line of sight and the corona's normal, and it varies over the precession period. The cosine of $\theta$ can be expressed using the following equation:
\begin{equation}\label{eq:theta}
\cos{\theta} = \sin{i}\cos{\Phi}\sin{\gamma}\cos{\omega}+\sin{i}\sin{\Phi}\sin{\gamma}\sin{\omega} + \cos{\gamma}\cos{i},
\end{equation}
where $i$ is the inclination angle between the line of sight of the observer and the black hole spin axis, $\Phi$ is the azimuth angle of the observer, $\gamma$ is the included angle between the corona's normal and the black hole spin axis and $\omega$ is the precession phase angle, which varies from 0 to 2$\pi$ during the precession period.
We set $i=63^\circ$, $\Phi=180^\circ$ (resulting in the peak flux occurring at 0.5 QPO cycles) and $\gamma=5^\circ$ to calculate modulations of the optical depth and plot the results in Figure~\ref{fig:7} as dashed black lines. It is evident that the theoretical curves closely match the data points calculated using Equation~(\ref{eq:tau}) across all four epochs. For the four distinct epochs, we adjust the values of $\tau_0$ and $h/r$ to make the calculated modulation curves comparable to the data points. However, we note that the initial parameter values solely influence the amplitude and mean values, without affecting the overall trend of the modulations in $\tau$. It is important to note that the simplified model mentioned above disregards the Doppler boosting effect. In reality, the modulation of the X-ray flux is primarily influenced by a combination of the projected area and Doppler boosting effects. However, maximal Doppler boosting occurs when the flow is viewed most edge-on, since blue shifts from the approaching material dominate over red shifts from receding material. On the other hand, the projected area is largest when the flow is viewed most face-on. Therefore, the X-ray flux as a function of precession phase represents the competition between these two effects \citep{2013ApJ...778..165V,2015ApJ...807...53I}. In Figure~\ref{fig:7}, we can observe a good alignment between the data points and the theoretical curves calculated using Equation (\ref{eq:9}), which considers only the projected area effect. This suggests that the modulation of the X-ray flux across the four epochs is likely dominated by the projected area effect.

%\subsubsection{Modulation of Reflection Fraction}
As shown in Figure~\ref{fig:6}, we detect significant modulations of reflection fraction ($R_{\rm f}$) in Epochs 3 and 4. Reflection fraction is defined as the ratio of Comptonized emission intensity that illuminates the disk to the Comptonized emission intensity that directly reaches the observer. This parameter is tied to the intrinsic accretion geometry \citep{2016A&A...590A..76D}. Therefore, the observed modulation in $R_{\rm f}$ reflects that the accretion geometry keeps changing over the cycle and thus the solid angle of the emitter as seen by the reflector and/or the solid angle of the reflector as seen by the observer will also change accordingly. Consequently, the observed modulations in reflection fraction provide strong support for the geometric origin of the QPO. For instance, within the framework of the corona's L-T precession model proposed by \citet{2009MNRAS.397L.101I}, it is extensively anticipated that the reflected emission would display variations throughout the QPO period \citep{2012MNRAS.427..934I, 2015MNRAS.446.3516I, 2016MNRAS.461.1967I, 2018ApJ...858...82Y,2020ApJ...897...27Y,2022MNRAS.511..255N}, subsequently leading to corresponding modulations in the derived reflection fraction \citep[see][]{2020ApJ...897...27Y}. 

\subsubsection{Jet Precession Model}
Another geometric model worth considering is the jet precession model proposed by \citet{2021NatAs...5...94M,2023ApJ...948..116M}. In this model, the main picture is that, the emission on-axis is significantly stronger than that off-axis due to the Doppler boosting, since typical speeds at the base of the jet can already be relativistic \citep{2006MNRAS.368.1561M}. As a result, the observed flux from the precessing jet can be modulated. In this case, the speed of the jet material can effectively impact the QPO variability. Specifically, the modulation amplitude increases with higher jet speeds. This is because larger speed leads to greater differences in emission strength between the on-axis and off-axis cases.

\begin{figure}
\centering
    \includegraphics[width=0.45\linewidth]{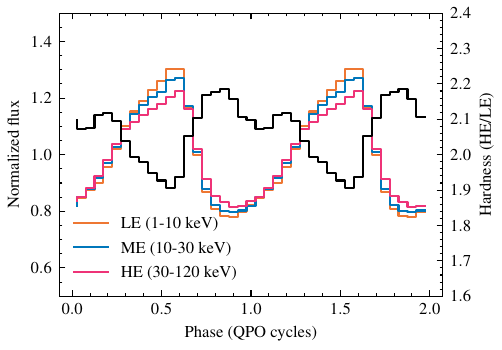}
    \caption{Constructed QPO waveforms of Epoch 1 by phase-folding the lightcurve from LE, ME and HE in the energy ranges of 1--10 keV (orange), 10--30 keV (blue) and 30--120 keV (pink), respectively. The black line represents the hardness defined as ratio between constructed QPO waveforms of HE and LE.}
    \label{fig:8}
\end{figure}

\begin{figure}
\centering
\begin{minipage}[c]{0.24\textwidth}
\centering
    \includegraphics[width=\linewidth]{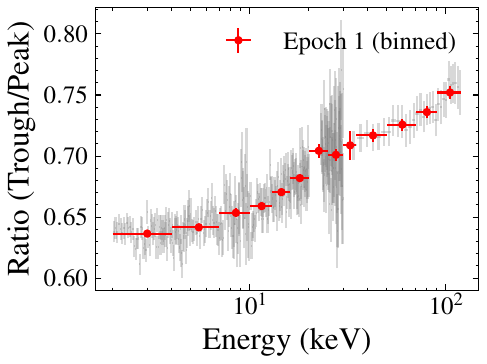}
\end{minipage}
\begin{minipage}[c]{0.24\textwidth}
\centering
    \includegraphics[width=\linewidth]{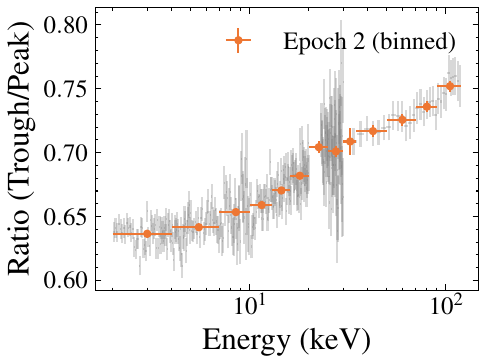}
\end{minipage}
\begin{minipage}[c]{0.24\textwidth}
\centering
    \includegraphics[width=\linewidth]{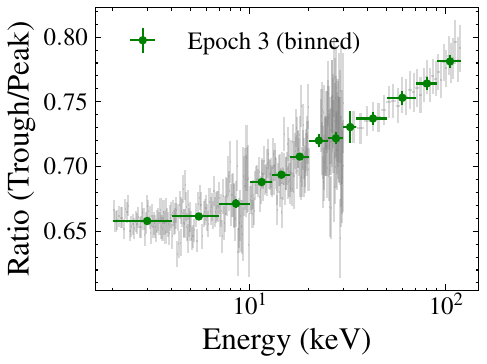}
\end{minipage}
\begin{minipage}[c]{0.24\textwidth}
\centering
    \includegraphics[width=\linewidth]{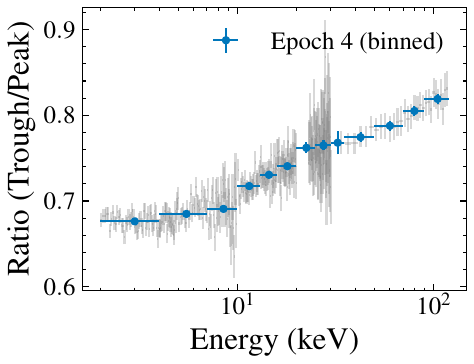}
\end{minipage}\\
\begin{minipage}[c]{0.24\textwidth}
\centering
    \includegraphics[width=\linewidth]{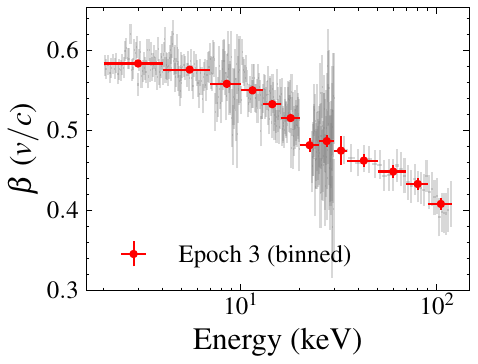}
\end{minipage}
\begin{minipage}[c]{0.24\textwidth}
\centering
    \includegraphics[width=\linewidth]{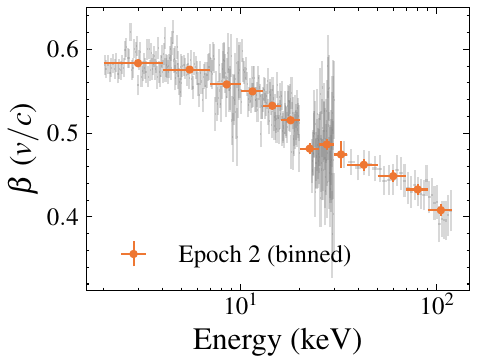}
\end{minipage}
\begin{minipage}[c]{0.24\textwidth}
\centering
    \includegraphics[width=\linewidth]{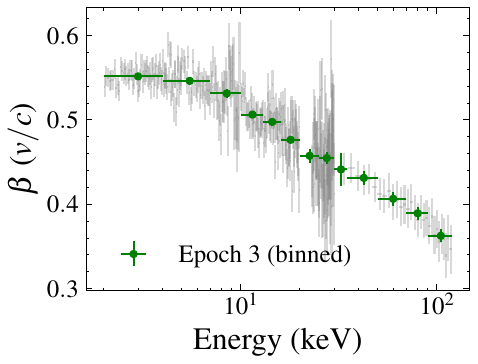}
\end{minipage}
\begin{minipage}[c]{0.24\textwidth}
\centering
    \includegraphics[width=\linewidth]{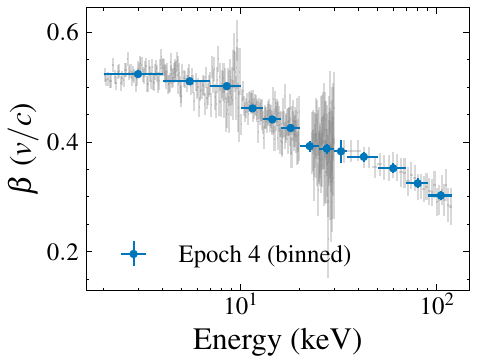}
\end{minipage}
\caption{\textit{Top}: the ratio between the spectra of QPO trough (0--0.1 cycles) and peak (0.4--0.5 cycles) phases across the four epochs. \textit{Bottom}: the energy dependence of the jet speed ($\beta$) across the four epochs, where values of $\beta$ are calculated by Equation~\ref{eq:15}. The gray scatters represent the calculation from the original spectra, while colored scatters represent rebinned values.}
    \label{fig:9}
\end{figure}

%\subsubsection{Modulation of Spectral Index}
\citet{2021NatAs...5...94M}, \citet{2022MNRAS.515.1914Z}, \citet{2023ApJ...948..116M} and \citet{2023arXiv230512317Y} have recently found that the soft energy photons intrinsically lag to the hard ones for the QPO signals in MAXI J1820+070. In the context of the jet precession model, these soft lags can be attributed to the differences among emitting regions for the emitted photons in different energy bands within the jet, as well as the geometric twist of the precessing jet itself. Typically, the harder photons are emitted at the lower part of the jet, while softer photons originate as the jet evolves further out. As the jet undergoes precession, different regions of the jet experience different precession phases, resulting in the observed soft lags. This original idea assumes that the speeds of the jet materials in different regions are the same. However, the jet at the base is at rest and from there on the particles are very efficiently accelerated \citep{2006MNRAS.368.1561M}. This means the jet does not have a uniform velocity \citep[see also][]{2013MNRAS.430.1694D}. If the QPO is indeed produced by the jet precession, the observed similarity in modulated phases between $\Gamma$ and the total flux suggests that the jet speed increases from lower to higher heights. As the speeds vary across different heights (corresponding to different energy bands of the emission), the modulation amplitude at specific phases differs for each energy band. Specifically, lower energy photons are emitted from higher-speed ejecta, resulting in the normalized flux of lower energies being larger than that of higher energies during the peak phases. During the trough phases, the situation is reversed, where the normalized flux of higher energies becomes larger than that of lower energies. Therefore, the hardness should vary over the precession cycle. In Figure~\ref{fig:8}, we plot the QPO waveforms for three instruments. To plot this figure, we perform the HHT to the light curves from each instrument and phase-fold them, then modify the phase differences among the three instruments. The plotted QPO waveforms are normalized by the averaged fluxes. Additionally, we plot the hardness, calculated as the ratio of the count rate in the HE band to that in the LE band within each phase intervals, in Figure~\ref{fig:8} as a black line. Clearly, the hardness exhibits an inverse modulation with respect to the flux, and the variability amplitude shows an increasing trend with energy, which is consistent with RMS spectrum presented by \citet{2021NatAs...5...94M}. However, it is important to note that the maximum soft lags (photons at $\sim100$ keV with respect to that at $\sim2$ keV) are $\sim$0.5 s at the QPO frequency of $\sim$0.04 Hz \citep{2021NatAs...5...94M}, indicating a phase lag of $<0.05$ QPO cycles. Consequently, the phase lags between different energy bands are not expected to have a significant influence on the phase-resolved spectral analysis. Therefore, the spectra obtained at the peak and trough phases in Section~\ref{sec:3} allow us to establish the relationship between the speed of the ejecta and the energy of the corresponding emitted photons. For a continuous jet, the observed flux from the jet can be written by
\begin{equation}
F_{\rm obs} = F_{\rm int}\mathcal{D}^{\Gamma+2},
\end{equation}
where $F_{\rm obs}$ is the observed spectrum, $F_{\rm int}$ is the intrinsic spectrum of the jet, $\Gamma$ is the spectral index of the jet emission and $\mathcal{D}$ is the relativistic boosting factor given by
\begin{equation}
\mathcal{D}=\frac{(1-\beta^2)^{1/2}}{1-\beta\cos{\theta}},
\end{equation}
where $\beta=v/c$ is the jet speed in units of the speed of light, and $\theta$ is the included angle between the jet axis and the line of sight of the observer, which is calculated by Equation~(\ref{eq:theta}). Therefore, the ratio of the trough spectrum to peak spectrum can be expressed as
\begin{equation}\label{eq:14}
\frac{F_{\rm t}}{F_{\rm p}}=\left[\frac{1-\beta\cos\theta_{\rm p}}{1-\beta\cos\theta_{\rm t}}\right]^{\Gamma+2},
\end{equation}
where $F_{\rm p}$ and $F_{\rm t}$ are the spectra at the peak and trough phases, respectively, and $\theta_{\rm p}$ and $\theta_{\rm t}$ are the included angles at the peak and trough phases, respectively. In the case of $\Phi=180^\circ$, $\theta_{\rm t}=\theta_{\omega=0}$ and $\theta_{\rm p}=\theta_{\omega=\pi}$, respectively. We define the ratio between the trough spectrum and peak spectrum as $\mathcal{R}=F_{\rm t}/F_{\rm p}$, which is a function of energy, and plot the calculated results for four epochs in the top panels of Figure~\ref{fig:9}. It can be observed that $\mathcal{R}$ exhibits an increasing trend with energy across four epochs, indicating that the trough spectra is harder than the peak spectra. From Equation~(\ref{eq:14}), it can be deduced that
\begin{equation}\label{eq:15}
\beta(E)=\left[\mathcal{R}^{1/(\Gamma+2)}-1\right]\cdot\left[\mathcal{R}^{1/(\Gamma+2)}\cos\theta_{\rm t}-\cos\theta_{\rm p}\right]^{-1},
\end{equation}
where $E$ is the energy of emitted photons. Here, we set $\Gamma$ as the value obtained from the spectral fitting of the phase-averaged spectra. The bottom panels of Figure~\ref{fig:9} displays the energy dependence of $\beta$. The obtained anti-correlation between $\beta$ and the energy indicates as the jet material is ejected away from the black hole and the energy of the emitted photons decreases, while the speed of the jet material increases. We also observe a slight decrease in the computed jet speed in each energy band from Epochs 1 to 4. Additionally, \citet{2021NatAs...5...94M} presented a decreasing trend of QPO RMS during the same stage. These findings collectively suggest that the outflowing materials decelerate during the bright hard state according to the jet precession model. However, the evolution of $R_{\rm f}$ indicates an acceleration of the outflowing plasma \citep{2021NatCo..12.1025Y}. This apparent contradiction may arise from a potential oversimplification within the jet precession model, specifically in maintaining a constant included angle  between the jet and black hole spin axes ($\gamma$). In fact, the coupling between the jet and inner disk \citep[see][]{2023ApJ...948..116M} could entail a decrease in $\gamma$ as the inner radius of the disk moves inwards. This behavior arises due to the tendency of the disk to align with the black hole spin, driven by the Bardeen-Petterson (BP) effect \citep{1975ApJ...195L..65B}. Consequently, the precession of the jet may occur with a reduced $\gamma$. Given that the QPO RMS is influenced by both the jet speed and $\gamma$, the combined impact of jet acceleration and the $\gamma$ reduction may account for the observed declining trend in QPO RMS as well as the computed jet speed using Eq. (\ref{eq:15}). It is important to note that the derived trend of $\beta$-$E$ correlation within each epoch remains unaffected by the assumption of a constant $\gamma$.

%It is important to note that Eq.~(\ref{eq:15}) is derived under the assumption that the observed spectrum is dominated by the emission from the jet, while the contributions from the disk, coronal and reflected components can be disregarded. However, the simplified formula represented by Eq.~(\ref{eq:15}) already provides a rough estimation of the energy dependence of the jet speed, although the refined $\beta$-$E$ relation needs more precise calculations.
%In context of the jet precession model, these three additional components can be likened to the direct-current (DC) part, while the jet emission represents the alternating-current (AC) part. By subtracting the spectra of the trough off the peak, we can isolate the pure AC component, which is related to the intrinsic jet emission. Analyzing these subtracted spectra allows for a more precise investigation of the jet properties, which will be explored in more detail in future works.

\begin{figure}
\centering
\begin{minipage}[c]{0.24\textwidth}
\centering
    \includegraphics[width=\linewidth]{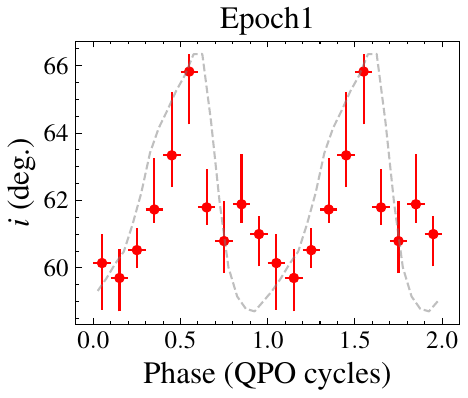}
\end{minipage}
\begin{minipage}[c]{0.24\textwidth}
\centering
    \includegraphics[width=\linewidth]{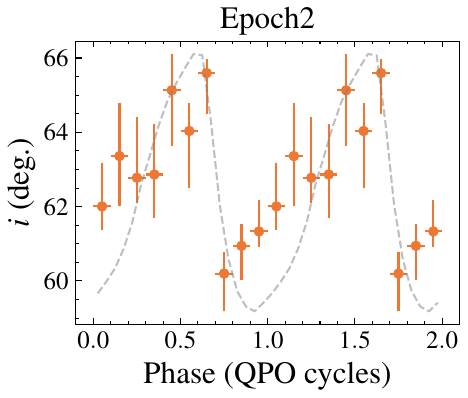}
\end{minipage}
\begin{minipage}[c]{0.24\textwidth}
\centering
    \includegraphics[width=\linewidth]{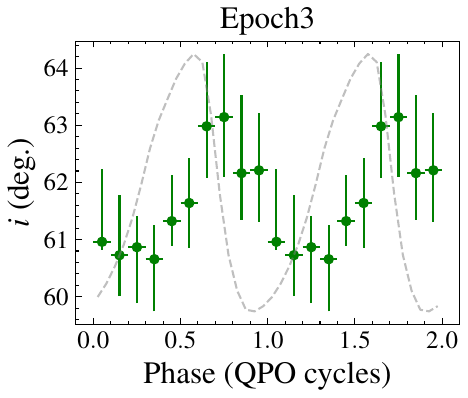}
\end{minipage}
\begin{minipage}[c]{0.24\textwidth}
\centering
    \includegraphics[width=\linewidth]{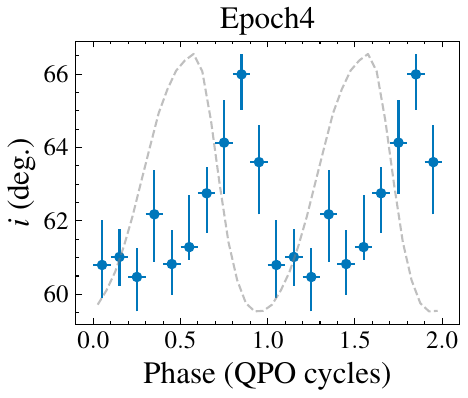}
\end{minipage}
\caption{Phase dependence of the inclination across the four epochs. To obtain the modulations in the inclination ($i$), we set reflection fraction to the best-fit values acquired from the phase-averaged spectral fittings. The uncertainties are reported at the 68\% confidence using the MCMC technique. Additionally, we include the QPO waveform in each panel as the gray dashed line.}
    \label{fig:10}
\end{figure}

%\subsubsection{Modulation of Reflection fraction}
Using Insight-HXMT and NICER observations, \citet{2023ApJ...948..116M} presented the energy dependence of the LFQPO characteristics and phase lags from 0.2 to 200 keV, which could be explained by the picture that the jet and inner disk ring precessing together. In this model, the inner region of the jet and the innermost part of the accretion disk are somehow coupled, resulting in a scenario where the precession of the disk induces a corresponding precession of the jet. %This picture is supported by a recent general relativistic magneto-hydrodynamic (GRMHD) simulation which demonstrated that the accretion flow and jets precess together, owing to the combination of Lense–Thirring effects and pressure or magnetic torques from the inflow/outflow system \citep{2018MNRAS.474L..81L,2023NewA..10102012L}.
Crucially, the geometrical alignment between the jet and the inner disk remains constant throughout the precession cycle. In this case, if the height of the X-ray emission region within the jet is sufficiently low, thereby solely illuminating the precessing inner disk, the reflection fraction ($R_{\rm f}$) should in principle remain constant over the QPO cycle. However, the inclination of the disk, denoted as $i$, is expected to undergo modulations that contrast with the flux variations. Taking this into account, we set $R_{\rm f}$ to the best-fit values acquired from the phase-averaged spectral fittings, while simultaneously allowing the inclination angle, $i$, to be a free parameter in the phase-resolved spectral fittings. Following the adjustment to spectral fittings, we also obtain a reasonable value of $\chi^2$ ($\sim$1), and compute the uncertainties of $i$ at the 90\% confidence using the MCMC technique. The behavior of $i$ across the QPO cycle for the four epochs under investigation is depicted in Figure~\ref{fig:10}. Clearly, the inclination exhibits notable modulations and shows large phase lags relative to the flux in Epochs 3 and 4, whereas positive correlations between $i$ and flux are observed in Epochs 1 and 2. This modulation pattern of $i$ during Epochs 3 and 4 lends support to the aforementioned jet-disk coupling precession model.
On another note, our spectral fitting model employs $xillverCp$ to accommodate the contribution arising from the distant reflector component. Intriguingly, the modulation of $xillverCp$ normalization seems to be more conspicuous during Epochs 1 and 2 in contrast to Epochs 3 and 4 (see Figure~\ref{fig:6}).
It is challenging to explain these distinctions across four epochs. One plausible interpretation is that the underlying phase-averaged accretion geometry undergoes changes during the bright hard state.

From a geometric perspective, a higher vertically extended jet would naturally have a larger solid angle with respect to the accretion disk, thereby efficiently illuminate the outer disk. As a result, both the inner precessing and the outer non-precessing regions of the disk could effectively receive incident flux originating from the vertically extended jet. However, it is important to note that modulations in the reflection component from these two regions are in principle different, owing to the fact that the inner disk undergoes precession to align with the jet, while the outer disk remains stable. Consequently, the resulting modulations in reflection fraction may potentially undergo some degree of smearing out. Additionally, the reflection model $relxillCp$ assumes a symmetric geometry \citep[see e.g.][]{2013MNRAS.430.1694D}, while the illumination pattern in the above precession scenario is asymmetric. Therefore, the positive correlation between $i$ and flux over the QPO cycle, as shown in Epochs 1 and 2, might be attributed to unclear parameter couplings within the current reflection models. The resolution of this matter warrants a detailed investigation through further simulations.
% due to parameter couplings within the spectral model. 

As mentioned above, if the height of the X-ray emission region within the jet is small enough to illuminate the inner precessing disk solely, modulations of $R_{\rm f}$ can be readily detected. Because the inner precessing disk is the primary reflector, leading to a more singular modulation pattern of reflection component. Furthermore, the observed substantial phase differences between the inclination angle ($i$) and the flux, align with the fundamental concept of the jet-disk coupling precession model. We propose that the picture of the decreasing jet emission height is consistent with the scenario of a contracting Comptonization region during the bright hard state \citep[][]{2019Natur.565..198K,2021NatCo..12.1025Y,2022ApJ...930...18W}. It is important to acknowledge that the jet-disk (where the disk corresponds to the geometrically thin and optically thick reflector) coupled precession model has not been supported by recent general relativistic magneto-hydrodynamic (GRMHD) simulations. For instance, \citet{2018MNRAS.474L..81L} demonstrated that the thick accretion flow and jets precess together, owing to the combination of Lense–Thirring effects and pressure or magnetic torques from the inflow/outflow system. However, their simulation did not include the optically thick disk. More recently, a simulation conducted by \citet{2023MNRAS.520L..79B} included both thin and thick accretion disks, revealing that the thick disk undergoes precession while the thin disk remains relatively stable. From our results, it remains uncertain whether the thin disk can undergo precession. Therefore, further investigation using numerical simulations and observations is necessary to explore this aspect in the future.

Regarding the modulation of the distant reflector component, it is not at odd that modulation strength decreases when the height of the jet becomes significantly smaller in the jet-disk coupling precession model. However, it is crucial to determine whether or not the modulation of Norm$_2$ is physical. Specifically, if the light-crossing time is longer than the QPO period, the modulation of the distant reflection component will be washed out. For the observed QPOs in Epoch 1, with a period of $\sim20$ s, the corresponding light-crossing distance is estimated to be $\sim6\times10^6$ km (equivalent to $\sim4.8\times10^5 R_{\rm g}$, assuming $M_{\rm BH}=8.5 M_{\odot}$). Based on the well determined orbital parameters of the system, the outer disc radius is constrained to be between $\sim6.3\times10^{4}$ and $\sim2.5\times10^{5} R_{\rm g}$ \citep{2023MNRAS.521.4190K}. Therefore, if the reflector is located within the distant accretion disk, it is possible to observe modulation in the distant reflection component. However, no significant phase lags between the total flux and this component, as shown in Figure~\ref{fig:6}, may suggest that the modulation of Norm$_2$ in Epochs 1 and 2 is not likely from reflection from the distant disk.

\section{Conclusion}\label{sec:5}
With the VMD algorithm and Hilbert transform applied to the LFQPOs of MAXI J1820+070 observed by Insight-HXMT, the QPO waveforms are constructed.
The phase-resolved energy spectra show strong phase modulation of the spectral index concurrent to the source flux. Similar phase modulations are found for the reflection fraction and evolve with outburst. These findings provide new insight to understanding the QPO in aspects of origination and evolution within outburst.  
%% IMPORTANT! The old "\acknowledgment" command has be depreciated. It was
%% not robust enough to handle our new dual anonymous review requirements and
%% thus been replaced with the acknowledgment environment. If you try to 
%% compile with \acknowledgment you will get an error print to the screen
%% and in the compiled pdf.
%% 
%% Also note that the akcnowlodgment environment does not support long amounts of text. If you have a lot of people and institutions to acknowledge, do not use this command. Instead, create a new \section{Acknowledgments}.
\begin{acknowledgments}
We are grateful to the anonymous referee for constructive comments that helped us improve this paper. This research has made use of data obtained from the High Energy Astrophysics Science Archive Research Center (HEASARC), provided by NASA’s Goddard Space Flight Center, and the Insight-HXMT mission, a project funded by China National Space Administration (CNSA) and the Chinese Academy of Sciences (CAS). This work is supported by the National Key R\&D Program of China (2021YFA0718500) and the National Natural Science Foundation of China under grants, U1838201, U1838202, 12173103, U2038101 and U1938103. This work is partially supported by International Partnership Program of Chinese Academy of Sciences (Grant No.113111KYSB20190020). Ling D. Kong is grateful for the financial support provided bu the Sino-German (CSC-DAAD) Postdoc Scholarship Program (57251553).
\end{acknowledgments}

%% To help institutions obtain information on the effectiveness of their 
%% telescopes the AAS Journals has created a group of keywords for telescope 
%% facilities.
%
%% Following the acknowledgments section, use the following syntax and the
%% \facility{} or \facilities{} macros to list the keywords of facilities used 
%% in the research for the paper.  Each keyword is check against the master 
%% list during copy editing.  Individual instruments can be provided in 
%% parentheses, after the keyword, but they are not verified.

%% Similar to \facility{}, there is the optional \software command to allow 
%% authors a place to specify which programs were used during the creation of 
%% the manuscript. Authors should list each code and include either a
%% citation or url to the code inside ()s when available.

%% Appendix material should be preceded with a single \appendix command.
%% There should be a \section command for each appendix. Mark appendix
%% subsections with the same markup you use in the main body of the paper.

%% Each Appendix (indicated with \section) will be lettered A, B, C, etc.
%% The equation counter will reset when it encounters the \appendix
%% command and will number appendix equations (A1), (A2), etc. The
%% Figure and Table counter will not reset.

\appendix

\section{MCMC Parameter Probability Distributions}
\label{appenix}
\begin{figure}
\centering
    \includegraphics[width=0.8\linewidth]{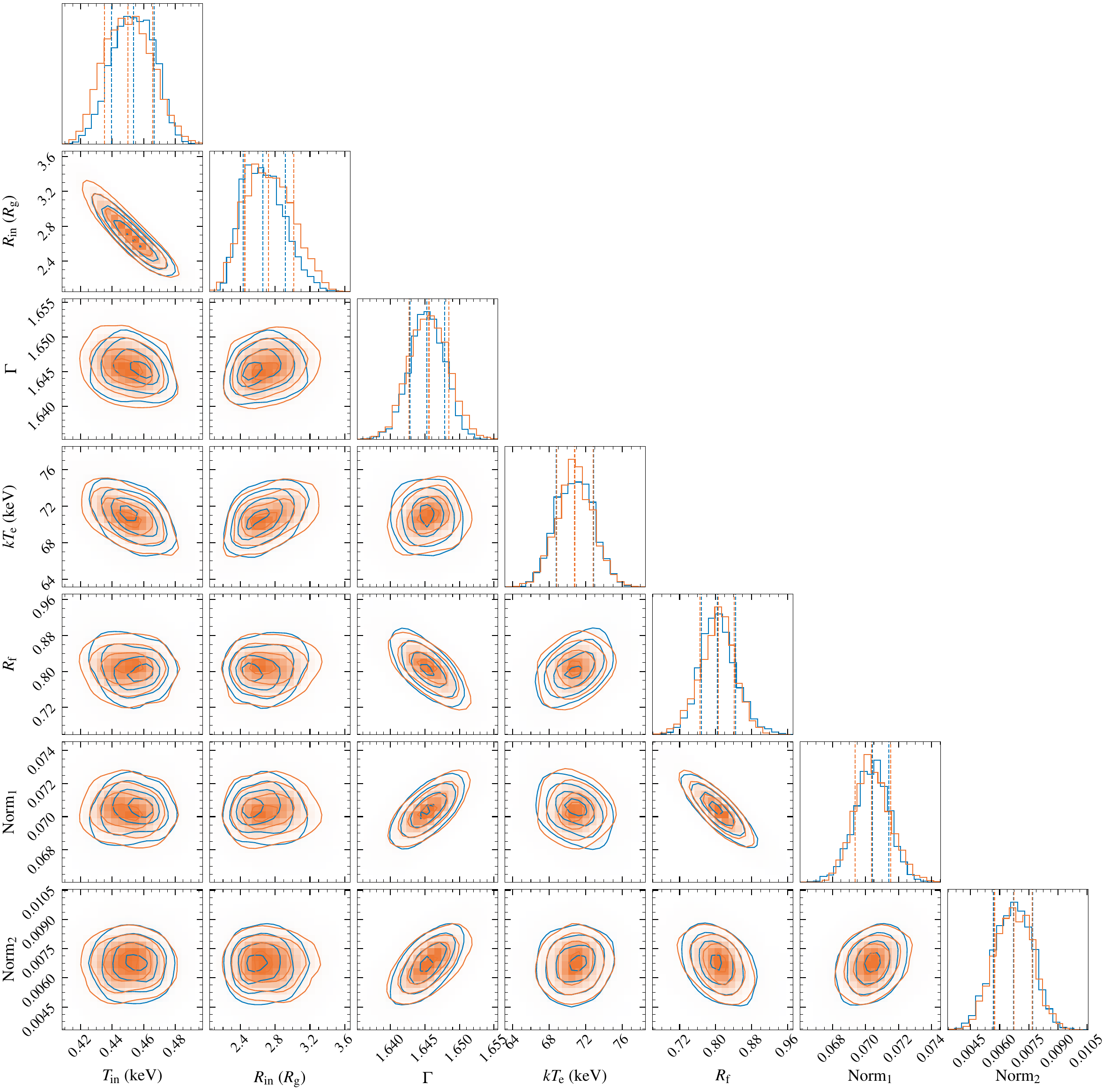}
    \caption{One- and two-dimensional projections of the posterior probability distributions, and the 0.16, 0.5 and 0.84 quantile contours derived from the MCMC analysis for each free spectral parameters. To test the convergence, we compare the one- and two-dimensional projections of the posterior distributions from the first (blue) and second (orange) halves of the chain, and we find no large differences. This illustration corresponds to the spectral fitting of the QPO trough phase in Epoch 4.}
    \label{fig:mcmc}
\end{figure}

In this section, we show an example (the QPO trough phase in Epoch 4) of the MCMC analysis for the spectral fitting model presented in Section~\ref{sec:3}. We use the Goodman-Weare algorithm with 8 walkers and a total length of 40,000 to perform the MCMC analysis, and the initial 2000 elements are discarded as “burn-in” period during which the chain reaches its stationary state. We find that the autocorrelation length is typically one thousand elements, so the net number of independent samples in the parameter space we have is order of $10^4$. Furthermore, we compare the one- and two-dimensional projections of the posterior distributions for each parameters from the first and second halves of the chain to test the convergence, and we find no significant differences (see Figure~\ref{fig:mcmc}). The contour maps and probability distributions are plotted using the \textit{corner} package \citep{2016JOSS....1...24F}.

%% For this sample we use BibTeX plus aasjournals.bst to generate the
%% the bibliography. The sample631.bib file was populated from ADS. To
%% get the citations to show in the compiled file do the following:
%%
%% pdflatex sample631.tex
%% bibtext sample631
%% pdflatex sample631.tex
%% pdflatex sample631.tex

\bibliography{sample631}{}
\bibliographystyle{aasjournal}

%% This command is needed to show the entire author+affiliation list when
%% the collaboration and author truncation commands are used.  It has to
%% go at the end of the manuscript.
%\allauthors

%% Include this line if you are using the \added, \replaced, \deleted
%% commands to see a summary list of all changes at the end of the article.
%\listofchanges

\end{document}